\documentclass[12pt,preprint]{aastex}
\def\degpoint{\ifmmode ^{\rm{o}}\!. \else $^{\rm{o}}\!.$\fi}
\newcommand{\degrees}{$^{\rm{o}}$}
\newcommand{\ms}{\mbox{m\,s$^{-1}$}}
\newcommand{\kms}{\mbox{km \ s$^{-1}$}}
\newcommand{\Msun}{\mbox{M$_{\odot}$}}

\newcommand{\Mjup}{\mbox{M$_{\rm Jup}$}}

\newcommand{\Mearth}{\mbox{M$_{\oplus}$}}

\newcommand{\ltsimeq}{\raisebox{-0.6ex}{$\,\stackrel
         {\raisebox{-.2ex}{$\textstyle <$}}{\sim}\,$}}

\usepackage{graphicx}

\begin{document}

\title{The Anglo-Australian Planet Search XXV: A Candidate Massive 
Saturn Analog Orbiting HD 30177 }

\author{Robert A.~Wittenmyer\altaffilmark{1,2}, Jonathan 
Horner\altaffilmark{1,2}, M.W. Mengel\altaffilmark{1}, R.P. 
Butler\altaffilmark{3}, D.J. Wright\altaffilmark{2}, 
C.G.~Tinney\altaffilmark{2}, B.D.~Carter\altaffilmark{1}, 
H.R.A.~Jones\altaffilmark{4}, G. Anglada-Escud\'e\altaffilmark{5}, 
J.~Bailey\altaffilmark{2}, Simon J.~O'Toole\altaffilmark{6} }

\altaffiltext{1}{Computational Engineering and Science Research Centre, 
University of Southern Queensland, Toowoomba, Queensland 4350, 
Australia}
\altaffiltext{2}{School of Physics and Australian Centre for 
Astrobiology, University of New South Wales, Sydney 2052, Australia}
\altaffiltext{3}{Department of Terrestrial Magnetism, Carnegie 
Institution of Washington, 5241 Broad Branch Road, NW, Washington, DC 
20015-1305, USA}
\altaffiltext{4}{Centre for Astrophysics Research, University of 
Hertfordshire, College Lane, Hatfield, Herts AL10 9AB, UK}
\altaffiltext{5}{School of Physics and Astronomy, Queen Mary University 
of London, 327 Mile End Road, London E1 4NS, UK}
\altaffiltext{6}{Australian Astronomical Observatory, PO Box 915,
North Ryde, NSW 1670, Australia}

\email{
rob@unsw.edu.au}

\shorttitle{HD 30177 Saturn Analog }
\shortauthors{Wittenmyer et al.}

\begin{abstract}

\noindent We report the discovery of a second long-period giant planet 
orbiting HD\,30177, a star previously known to host a massive Jupiter 
analog (HD\,30177b: a=3.8$\pm$0.1 au, m sin~$i=9.7\pm$0.5\,\Mjup).  
HD\,30177c can be regarded as a massive Saturn analog in this system, 
with a=9.9$\pm$1.0 au and m sin~$i=7.6\pm$3.1\,\Mjup.  The formal best 
fit solution slightly favours a closer-in planet at $a\sim$7\,au, but 
detailed \textit{n}-body dynamical simulations show that configuration 
to be unstable.  A shallow local minimum of longer-period, 
lower-eccentricity solutions was found to be dynamically stable, and 
hence we adopt the longer period in this work.  The proposed $\sim$32 
year orbit remains incomplete; further monitoring of this and other 
stars is necessary to reveal the population of distant gas giant 
planets with orbital separations $a\sim$10\,au, analogous to that of 
Saturn.

\end{abstract}

\keywords{planetary systems --- techniques: radial velocities -- stars: 
individual (HD 30177) }

\section{Introduction}

Prior to the dawn of the exoplanet era, it was thought that planetary 
systems around other stars would likely resemble our own - with small, 
rocky planets close to their host stars, and the more massive, giant 
planets at greater distances.  With the discovery of the first 
exoplanets, however, that paradigm was shattered - and it rapidly became 
clear that many planetary systems are dramatically different to our own.  
But to truly understand how unusual (or typical) the Solar system is, we 
must find true Jupiter and Saturn analogs: massive planets on 
decade-long orbits around their hosts.  The only way to find such 
planets is to monitor stars on decade-long timescales, searching for the 
telltale motion that might reveal such distant neighbours.

Nearly three decades of planet search have resulted in a great 
unveiling, at every stage of which we are finding our expectations 
consistently upturned as the true diversity of worlds becomes ever more 
apparent.  Much progress has been made in understanding the occurrence 
rates and properties of planets orbiting within $\sim$1 au of their 
stars, brought on by the \textit{Kepler} revolution 
\citep[e.g.][]{borucki11, rowe14, coughlin16} and the advent of Doppler 
velocimetry at precisions of 1\,\ms\ \citep{fischer16}.  While 
\textit{Kepler} has been hugely successful in exploring the frequency of 
planets close to their stars, such transit surveys are not suited to 
search for planetary systems like our own - with giant planets moving on 
orbits that take decades to complete.  To understand the occurrence of 
such systems requires a different approach - radial velocity monitoring 
of individual stars on decadal timescales.  

Sometimes overshadowed by the \textit{Kepler} dicoveries, but equally 
important for a complete picture of planetary system properties, are the 
results from ongoing ``legacy'' Doppler surveys, which are now sensitive 
to giant planets in orbits approaching 20 years.  Those surveys include, 
for example, the McDonald Observatory Planet Search 
\citep{texas1,endl16}, the California Planet Search \citep{howard10, 
feng15}, the Anglo-Australian Planet Search \citep{tinney01, jupiters, 
gj832}, and the Geneva Planet Search \citep{marmier13, moutou15}.

The emerging picture is that the Solar System is not typical of 
planetary systems in the Solar neighbourhood.  For example, 
super-Earths, planets with masses $\sim$3-10\,\Mearth, are extremely 
common yet are completely absent from our Solar System.  Jupiter-like 
planets in Jupiter-like orbits appear to be relatively uncommon, 
orbiting only about 6\% of solar-type stars \citep{newjupiters}.

Such a low incidence of true Solar system analogs is of particular 
interest in the context of astrobiology, and the search for truly 
Earth-like planets beyond the Solar system.  In the Solar System, 
Jupiter has played a key role in the formation and evolution of the 
planetary system - variously corralling, sculpting and destabilising the 
system's smaller bodies (and thereby likely contributing significantly 
to the introduction of volatiles, including water, to the early Earth).  
Over the system's more recent history, Jupiter has managed the flux of 
smaller objects towards the Earth, influencing (but not necessarily 
reducing) the frequency of impacts on the terrestrial planets.  It has 
long been argued that the presence of a true Jupiter analog would be an 
important selection factor for an Earth-like planet to be truly 
habitable - although many recent studies have suggested that this might 
not be the case \citep[e.g.][]{hj08, hjc10, hj10b, lewis13}.

The Anglo-Australian Planet Search (AAPS) monitored $\sim$250 solar-type 
stars for 14 years.  Of these, a subset of $\sim$120 stars continued to 
be observed for a further three years, with the primary aim of detecting 
Jupiter-mass planets in orbits $P>10$yr \citep{newjupiters}.  The AAPS 
has delivered a consistent 3\,\ms\ velocity precision since its 
inception, enabling the discovery of several Jupiter analogs 
\citep[e.g.][]{jones10, 142paper, 2jupiters}.
 
This paper is organised as follows: Section 2 outlines the AAT and HARPS 
observations of HD\,30177 and gives the stellar parameters.  Section 3 
describes the orbit-fitting procedures and gives the resulting planetary 
parameters for the HD\,30177 system.  In Section 4 we perform a detailed 
dynamical stability analysis of this system of massive planets.  Then we 
give our conclusions in Section 5.

\section{Observational Data and Stellar Properties}

HD\,30177 is an old, Sun-like star, with a mass within 5\% of Solar.  It 
lies approximately 54.7 parsecs from the Sun, and has approximately 
twice Solar metallicity.  The stellar parameters for HD\,30177 can be 
found in Table~\ref{stellarparams}.  We have observed HD\,30177 since 
the inception of the AAPS, gathering a total of 43 epochs spanning 17 
years (Table~\ref{AATvels}).  Precise radial velocities are derived 
using the standard iodine-cell technique to calibrate the instrumental 
point-spread function \citep{val:95,BuMaWi96}.  The velocities have a 
mean internal uncertainty of 3.9$\pm$1.2\,\ms.

HD\,30177 has also been observed with the HARPS spectrograph on the ESO 
3.6m telescope in La Silla.  At this writing, 20 epochs spanning 11 
years are publicly available at the ESO Archive.  Velocities were 
derived using the HARPS-TERRA technique \citep{ang12}, and are given in 
Table~\ref{harpsvels}.

\section{Orbit Fitting and Results}

The inner planet, HD\,30177b was first announced in \citet{tinney03}, 
with a relatively unconstrained period of 1620$\pm$800 days and 
m~sin~$i=7.7\pm$1.5 \Mjup.  Its orbit was updated in \citet{butler06} on 
the basis of observations that clearly spanned one full orbital period, 
to $P=2770\pm$100 days and m~sin~$i=10.5\pm$0.9 \Mjup.  We now present a 
further 10 years of AAT data, along with 11 years of concurrent HARPS 
data, to refine the orbit of this planet.  As a result of this 
additional data, we now find that the single-planet fit exhibits 
significant residuals, suggesting the presence of a second, very 
long-period object in this system.  We have added 6\,\ms\ of jitter in 
quadrature to both data sets; this brings the reduced $\chi^2$ close to 
1 for two-planet models.  A single-planet model now has a 
reduced $\chi^2$ of 7.1 and an rms of 17.3\,\ms.  As in our previous 
work \citep[e.g.][]{tinney11, songhu, BDBlah, ppps3}, we have used a 
genetic algorithm to explore the parameter space for the outer planet, 
fitting a simultaneous two-Keplerian model that allows the outer planet 
to take on periods 4000-8000 days and eccentricities $e<0.3$.  The best 
fit from the genetic algorithm results was then used as a starting point 
for a two-Keplerian fit (downhill simplex minimisation) within the 
\textit{Systemic Console} \citep{mes09}.

\begin{figure}
\plotone{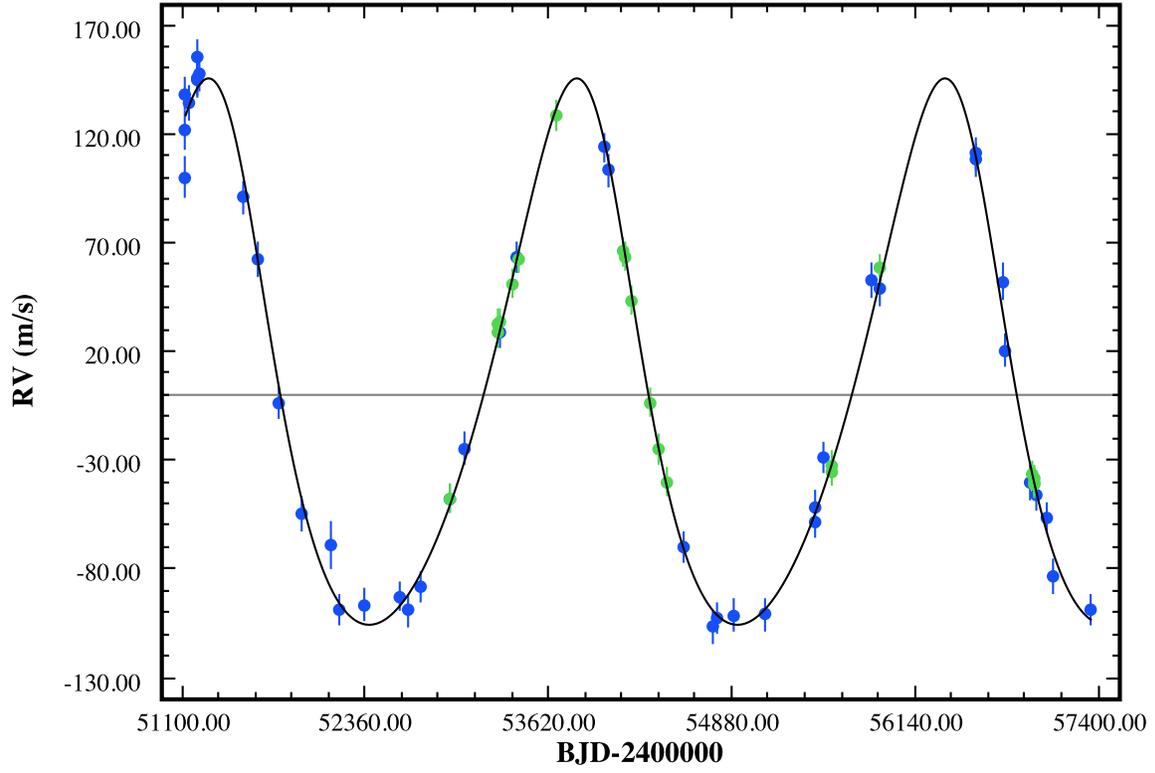}
\caption{Data and Keplerian model fit for the inner planet HD\,30177b.  
AAT -- blue; HARPS -- green.  The orbit of the outer planet has been 
removed.  We have added 6\,\ms\ of jitter in quadrature to the 
uncertainties, and this fit has an rms of 7.07\,\ms. }
\label{inner}
\end{figure}

\begin{figure}
\plotone{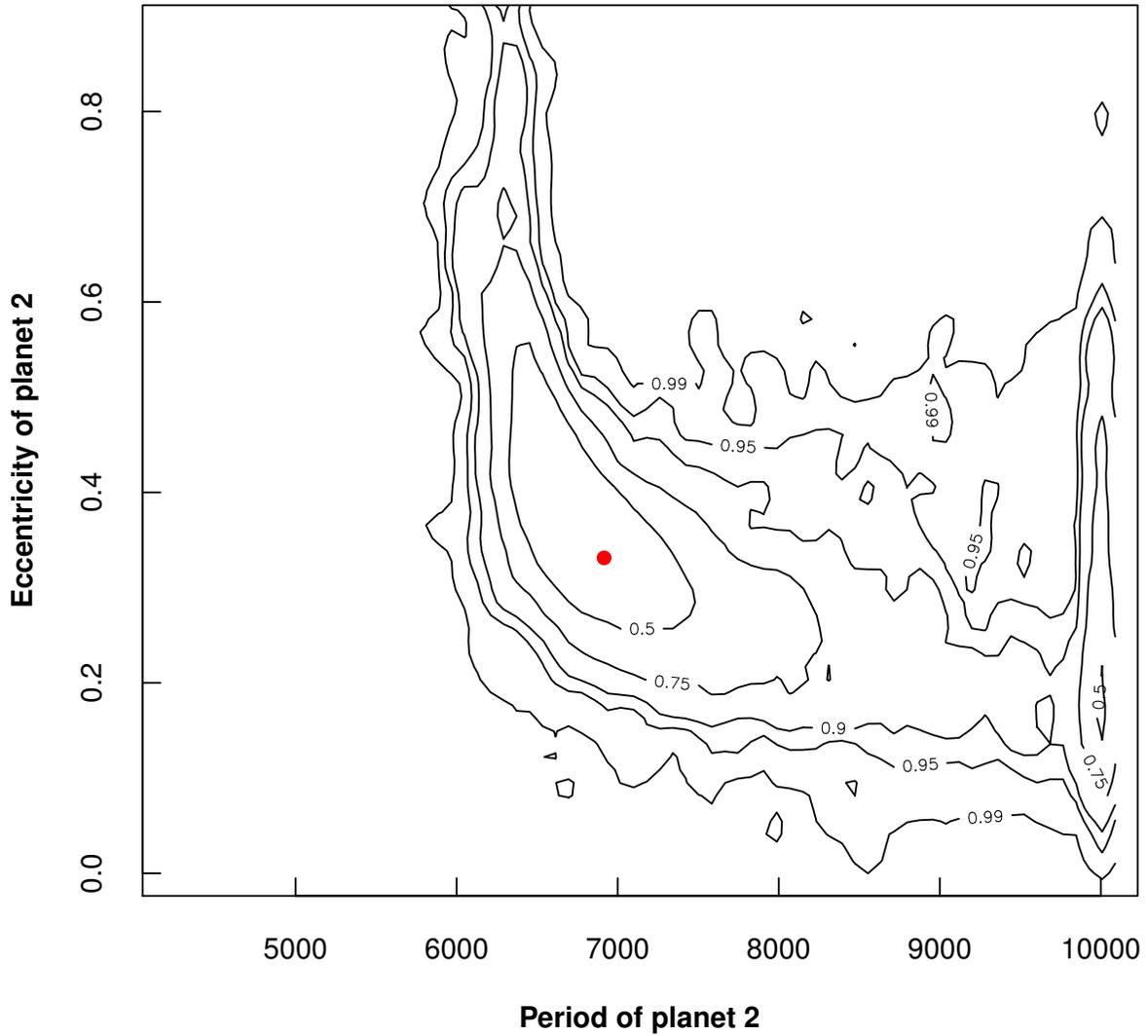}
\caption{Contours of $\chi^2$ as a function of the outer planet's 
eccentricity and orbital period.  Contours are labeled with confidence 
intervals around the best fit (red dot).  Hints of a second local 
$\chi^2$ minimum can be seen in the lower right, at long periods and low 
eccentricities. }
\label{chi1}
\end{figure}

\begin{figure}
\plottwo{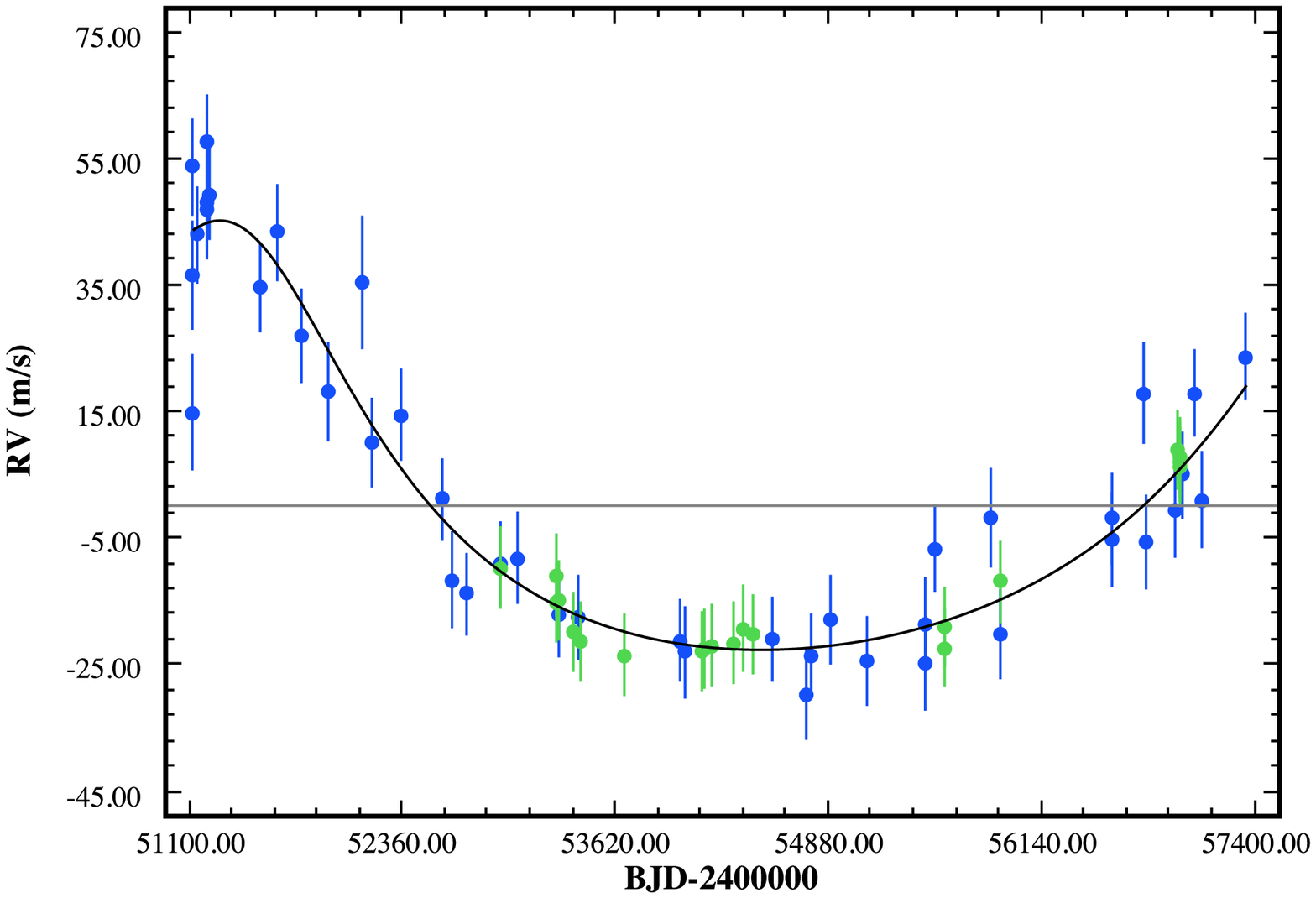}{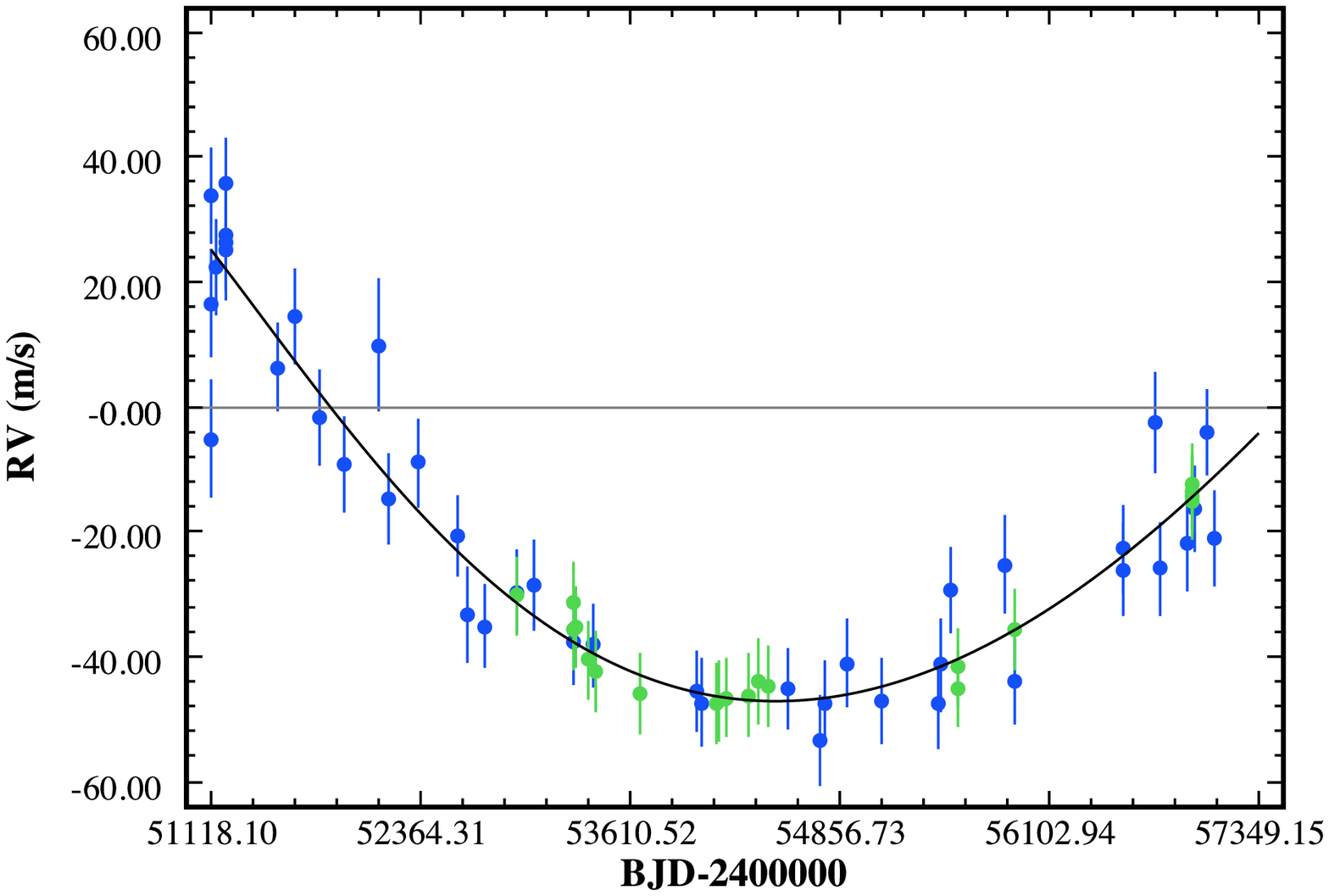}
\caption{Data and Keplerian model fit for the outer planet HD\,30177c. 
AAT -- blue; HARPS -- green.  The orbit of the inner planet has been 
removed.  We have added 6\,\ms\ of jitter in quadrature to the 
uncertainties.  Left panel: Nominal best fit, with $P=6921$ d.  Right 
panel: Long-period solution, with $P=11640$ d. }
\label{outerplanet1}
\end{figure}

The results of these fits are given in Table~\ref{planetparams1}.  The 
precision with which the parameters for the inner planet are known are 
now improved by a factor of ten, or more, over the previously published 
values \citep{butler06}.  The model fit for the inner planet is shown in 
Figure~\ref{inner}.  The nominal best fit solution features a second 
planet, HD\,30177c, with period of 6921$\pm$621 days and 
m~sin~$i=3.0\pm$0.3\,\Mjup.  We present both a ``best fit'' and a ``long 
period'' solution in recognition of the fact that for an incomplete 
orbit, the period can be wildly unconstrained and allow for solutions 
with ever-longer periods by adjusting the eccentricity.  
Figure~\ref{chi1} shows the $\chi^2$ contours as a function of the outer 
planet's period and eccentricity, based on the results from the best-fit 
solution given in the left columns of Table~\ref{planetparams1}.  The 
best fit solution appears to be a shallow minimum in the $\chi^2$ space, 
with a secondary minimum at lower eccentricity and longer period 
($P\sim$10000 days).  We thus attempted a second fit, starting the 
orbital period of the outer planet at 10000 days to guide the 
\textit{Systemic} simplex algorithm into the apparent secondary $\chi^2$ 
minimum seen in Figure~\ref{chi1}.  The results are given in the right 
columns of Table~\ref{planetparams1}, labelled ``long period.'' This fit 
results in an outer planet with period 11640$\pm$2432 days and 
m~sin~$i=6.4\pm$3.3\,\Mjup; the uncertainties are of course much larger 
since the available data only cover $\sim$60\% of the orbital period.  
The best-fit and long-period solutions are plotted in 
Figure~\ref{outerplanet1}.

\begin{figure}
\plotone{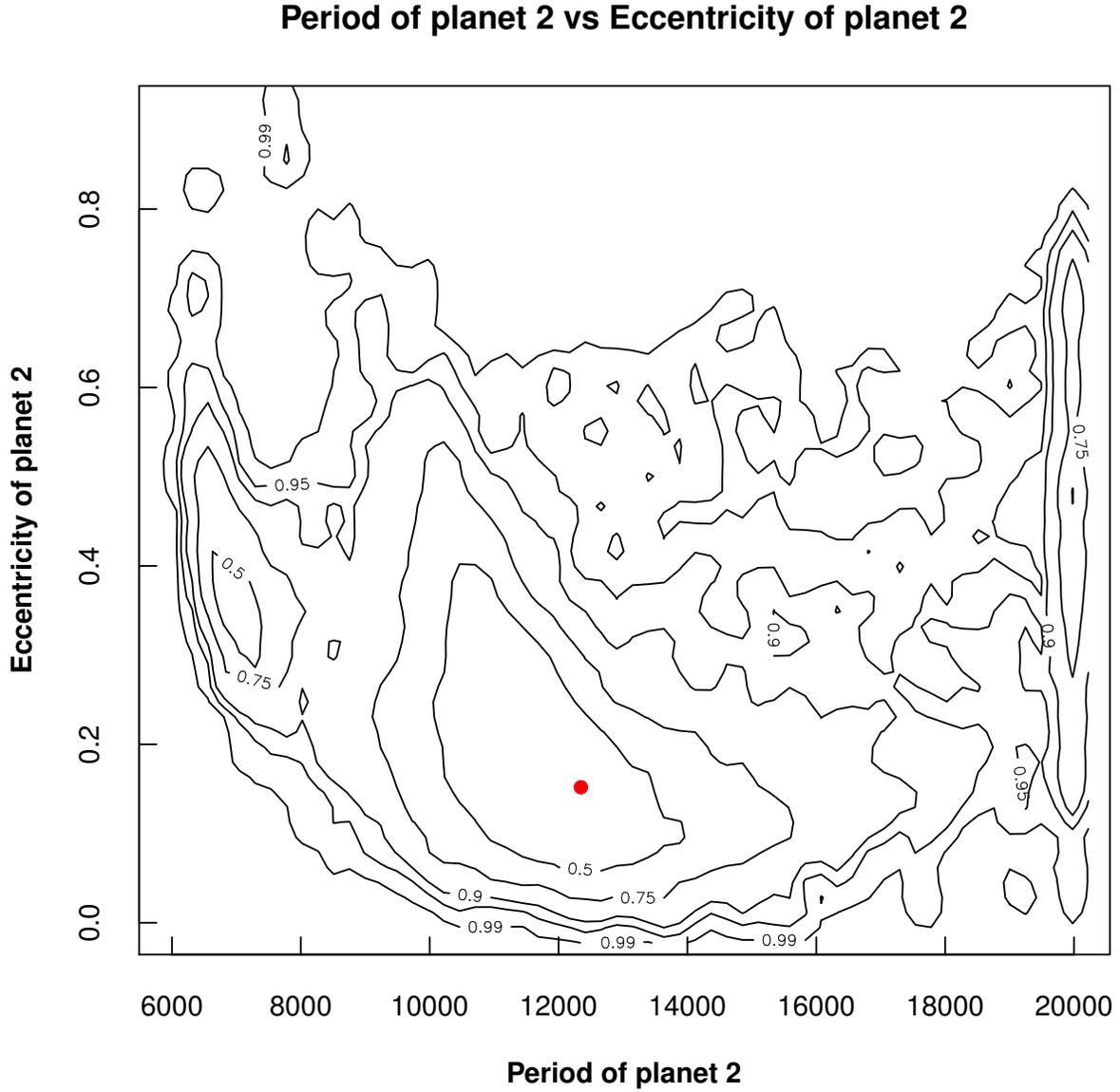}
\caption{Same as Figure~\ref{chi1}, but for the long-period solution 
where one outlier data point has been removed.  Two local $\chi^2$ 
minima are evident, with the longer-period solution at lower 
eccentricity (red dot). }
\label{chi2}
\end{figure}

One might argue that the outer planet hypothesis relies heavily on the 
presumption of a velocity turnover in the first few epochs, in 
particular the point at BJD 2451119, which lies about 30\,\ms\ below the 
previous night's velocity.  To check the effect of this potentially bad 
observation, we repeated the orbit fitting described above after 
removing that point.  The results are given in 
Table~\ref{planetparams2}, again expressed as a ``best fit'' and a 
``long period'' solution.  We now find a best fit at a period of 
7601$\pm$1134 days and m~sin~$i=3.3\pm$0.5\,\Mjup.  Removing the 
suspected outlier resulted in a slightly longer period that remains in 
formal agreement with the original solution in 
Table~\ref{planetparams1}.  For the long-period solution, we again 
started the \textit{Systemic} fitting routine at a period of 10,000 days 
for the outer planet.  This results again in a long period consistent 
with the long period solution obtained from the full set of velocities: 
we obtain a period of 11613$\pm$1837 days and 
m~sin~$i=7.6\pm$3.1\,\Mjup.  We thus have two possible solutions for the 
HD\,30177 two-planet system, which are virtually indistinguishable in 
terms of the RMS about the model fit or the $\chi^2$, due to the shallow 
minima and complex $\chi^2$ space (Figure~\ref{chi2}).  

\begin{figure}
\plotone{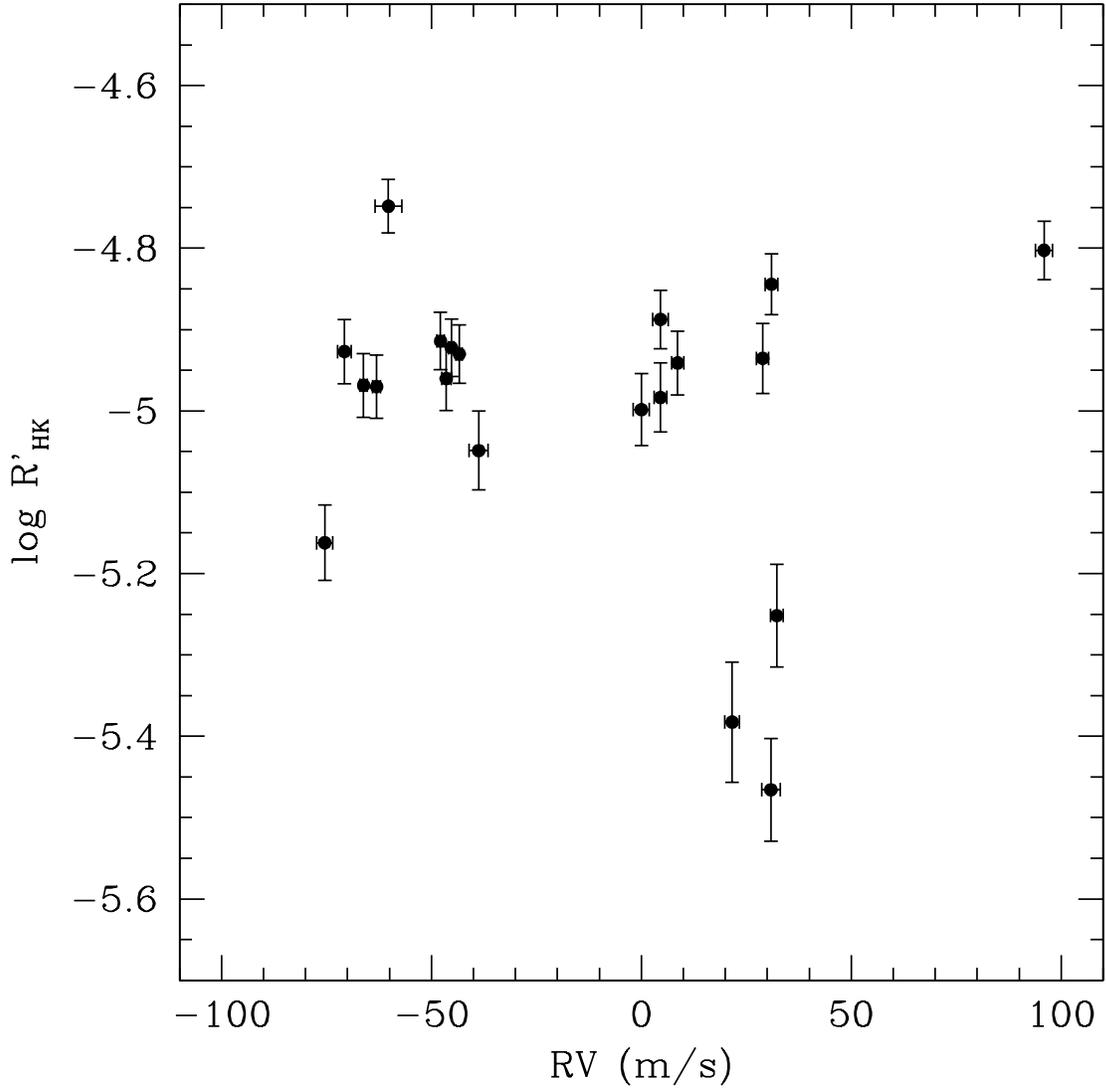}
\caption{Ca II activity index log$R'_{HK}$ as a function of radial 
velocity for the HARPS spectra of HD\,30177.  No correlation is evident 
from the 11 years of data, and hence we conclude that a stellar magnetic 
cycle is not responsible for the observed radial velocity variations. }
\label{calcium}
\end{figure}

For the old, solar-type stars generally targeted by radial 
velocity surveys, stellar magnetic cycles like the Sun's 11-year cycle 
are a concern when claiming detection of planets with orbital periods 
$\sim$10 years and longer.  While our AAT/UCLES spectra do not include 
the Ca II H and K lines most commonly used as activity proxies, the 
HARPS spectra used in this work do \citep[e.g.][]{dum11, lovis11, 
hebrard14}.  Figure~\ref{calcium} shows the Ca II activity log$R'_{HK}$ 
versus the HARPS radial velocities.  No correlation is evident.  
Clearly, a long-period body is present, but a longer time baseline is 
necessary to observe a complete orbit and better constrain its true 
nature.  In the next section, we explore the dynamical stability of the 
two candidate orbital solutions.


\section{Dynamical Stability Simulations}

In order to understand the dynamical context of the two distinct orbital 
solutions presented above, and to see whether they yield planetary 
systems that are dynamically feasible, we followed a now 
well-established route \citep[e.g.][]{MarshallHR8799,HD204313,QSVir}. 
For each solution, we performed 126,075 unique integrations of the 
system using the Hybrid integrator within the \textit{n}-body dynamics 
package, \textsc{Mercury} \citep{Chambers99}.  In each of those 
simulations, we held the initial orbit of the innermost planet fixed at 
its nominal best-fit values (as detailed in Table~\ref{planetparams1}). 
We then proceeded to systematically move the orbit of the outermost 
planet through the full $\pm 3\sigma$ uncertainty ranges for semi-major 
axis, $a$, eccentricity, $e$, argument of periastron, $\omega$, and mean 
anomaly, $M$.  In this manner, we created a regular grid of solutions, 
testing 41 unique values of $a$ and $e$, 15 unique values of $\omega$, 
and 5 unique values of $M$.

These simulations make two assumptions: first, that the two 
planets move on coplanar orbits (as is essentially the case in the Solar 
system), and second, we assign the planets their minimum masses 
(m~sin~$i$) as derived from the radial velocity data.  In a number of 
previous studies \citep[e.g.][]{HUAqr, BDBlah, hinse14}, we have 
examined the impact of mutual inclination on system stability.  However, 
for widely separated planets, the inclination between the two orbits 
seems to play little role in their stability.  It seems most likely that 
there would not be large mutual inclination between the orbits of the 
planets; from a dynamical point of view, given the assumption that the 
two planets formed in a dynamically cool disk, it is challenging to 
imagine how they could achieve significant mutual inclination without 
invoking the presence of a highly inclined distant perturber (i.e. an 
undetected binary companion, driving excitation through the Kozai 
mechanism).  It is certainly reasonable to assume that the orbits are 
most likely relatively coplanar, as is seen in the Solar system giant 
planets, and also in those multiple exoplanet systems with orbital 
inclinations constrained by transits \citep{fang12} or by resolved 
debris disk observations \citep{kennedy13}. 

Regarding the use of minimum masses, one would expect increased 
planetary masses to destabilise the systems.  The reason for this can be 
seen when one considers the ``gravitational reach'' of a planet, which 
can be defined in terms of its Hill radius. The Hill radius is 
proportional to the semi-major axis of the planet's orbit, but only 
increases as the cube root of the planet's mass.  As a result, doubling 
the mass of a planet only increases its gravitational reach, and 
therefore its Hill radius, by a factor of $2^(1/3) = 1.26$ - a 
relatively minor change.

The simulations were set to run for a maximum of 100 million years, but 
were brought to a premature end if one or other of the planets were 
ejected from the system or collided with the central body.  Simulations 
were also curtailed if the two planets collided with one another.  For 
each of these conditions, the time at which the simulation finished was 
recorded, allowing us to create dynamical maps of the system to examine 
the dynamical context of the orbits presented above, and to see whether 
the system was dynamically feasible.  Such maps have, in the past, 
revealed certain systems to be dynamically unfeasible 
\citep[e.g.][]{HUAqr,HUAqr2,QSVir,BDBlah}.  In other cases, dynamical 
mappings have resulted in stronger constraints for a given system's 
orbits than was possible solely on the grounds of the observational data 
\citep[e.g.][]{24Sex,texas1, HD73526}.  Dynamical simulations therefore 
offer the potential to help distinguish between different solutions with 
similar goodness of fit, such as those proposed in this work, as well as 
yielding an important dynamical 'sanity check'.

To complement these dynamical simulations, we also chose to trial a new 
technique for the dynamical analysis of newly discovered systems.  
Rather than populate regular grids in element space, whilst holding the 
better constrained planet's initial orbit fixed, we instead performed 
repeated fits to the observational data.  In our fitting, we required 
solely that the solutions produced lie within $3 \sigma$ of the best-fit 
solution, allowing all parameters to vary freely.  This created clouds 
of 'potential solutions' distributed around the best fit out to a range 
of $\chi^2_{best} + 9$.  We then randomly selected solutions to evenly 
sample the phase space between the best-fit solution (at 
$\chi^2_{best}$) and $\chi^2_{best}+9$.  As before, we generated 126,075 
unique solutions for each of the two scenarios presented above.

Where our traditional dynamical maps explore the dynamical context of 
the solutions in a readily apparent fashion, these new simulations are 
designed to instead examine the stability of the system as a function of 
the goodness of fit of the orbital solution.  In addition, they 
allow us to explore the stability as a function of the masses assigned 
to the two planets in question.  As such, they provide a natural 
complement to the traditional maps, as can be seen below.

\section{Dynamical Stability Results} 

Figure~\ref{ShortPeriodOrbit} shows the dynamical context of the short 
period solution for the two planet HD\,30177 system, as described in 
Table~\ref{planetparams1}.  The best fit solution lies in an area of 
strong dynamical instability, with the majority of locations within the 
1$\sigma$ uncertainty range being similarly unstable.  There is, 
however, a small subset of solutions in that range that are stable, 
marking the inner edge of a broad stable region seen towards larger 
semi-major axes and smaller eccentricities.  The small island of 
stability at $a \sim 5.687$~au is the result of the planets becoming 
trapped in mutual 2:1 mean-motion resonance, whilst the narrow curved 
region of moderate stability at high eccentricities is caused by orbits 
for HD\,30177c with periastron located at the semi-major axis of 
HD\,30177b.  Dynamical stability for the system on near-circular, 
non-resonant orbits is only seen in these simulations exterior to the 
planet's mutual 3:1 mean-motion resonance, located at $a \sim 7.453$~au 
(and the cause of the small island of stability at non-zero 
eccentricities at that semi-major axis).  As a result, these simulations 
suggest that the short-period solution is not dynamically favoured, 
unless the orbital period for HD\,30177c is significantly longer than 
the best fit, the orbit markedly less eccentric, or if the two planets 
are trapped in mutual 3:1 mean motion resonance.

\begin{figure}[SP_eccentricity]
\centering
  \begin{tabular}{@{}cc@{}}
    \includegraphics[width=.46\textwidth]{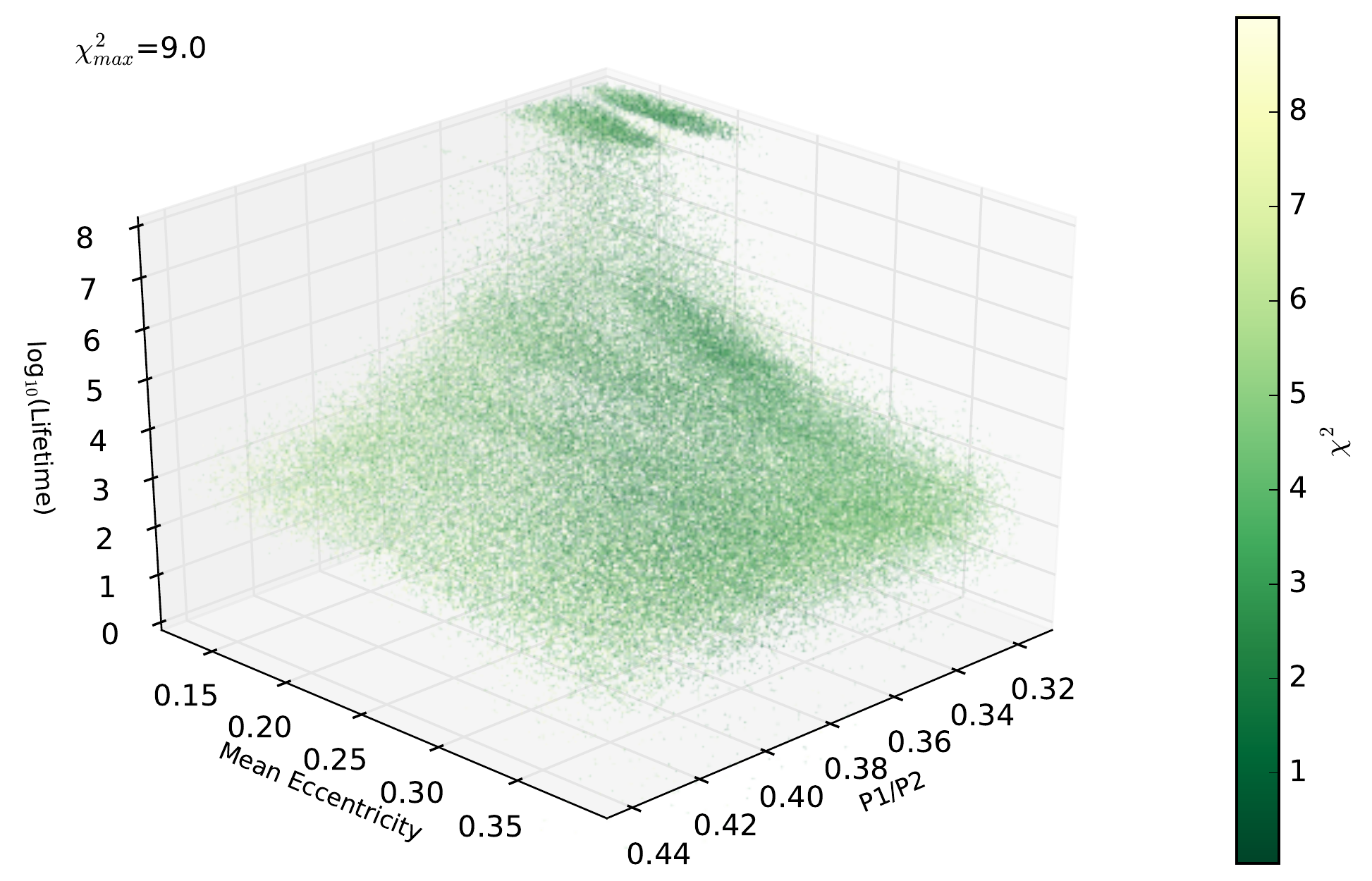} &
    \includegraphics[width=.46\textwidth]{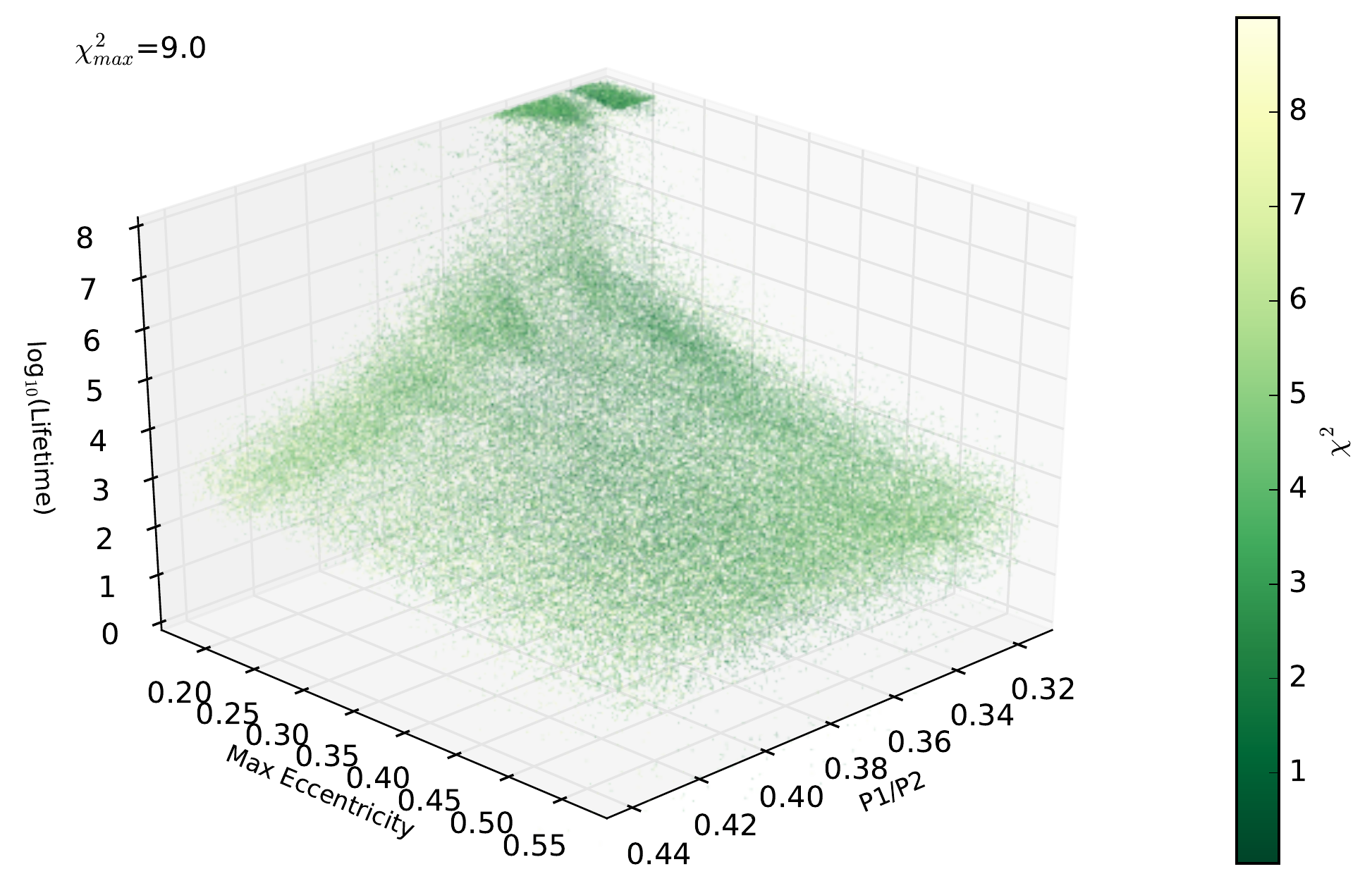} \\
    \includegraphics[width=.46\textwidth]{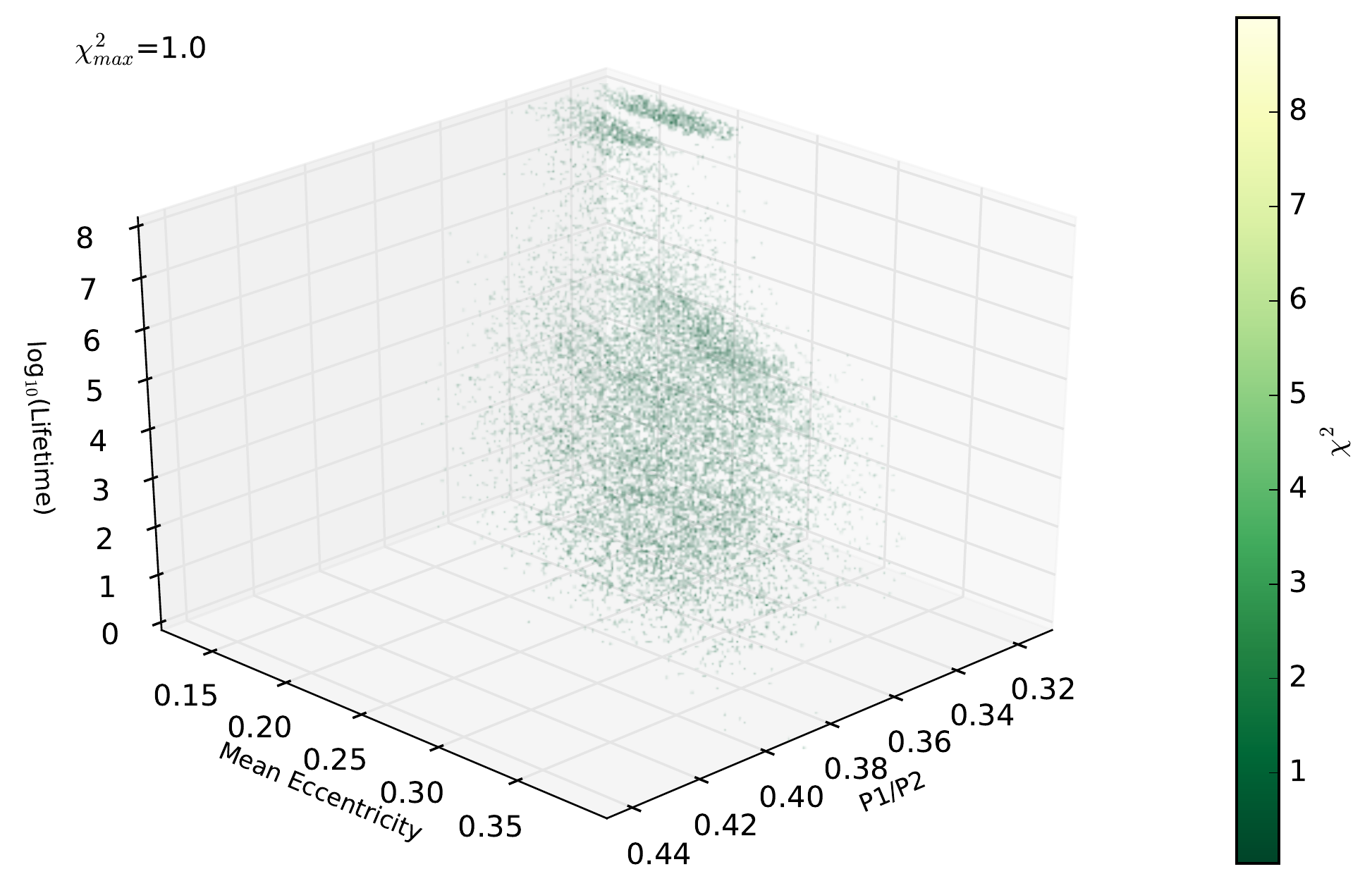} &
    \includegraphics[width=.46\textwidth]{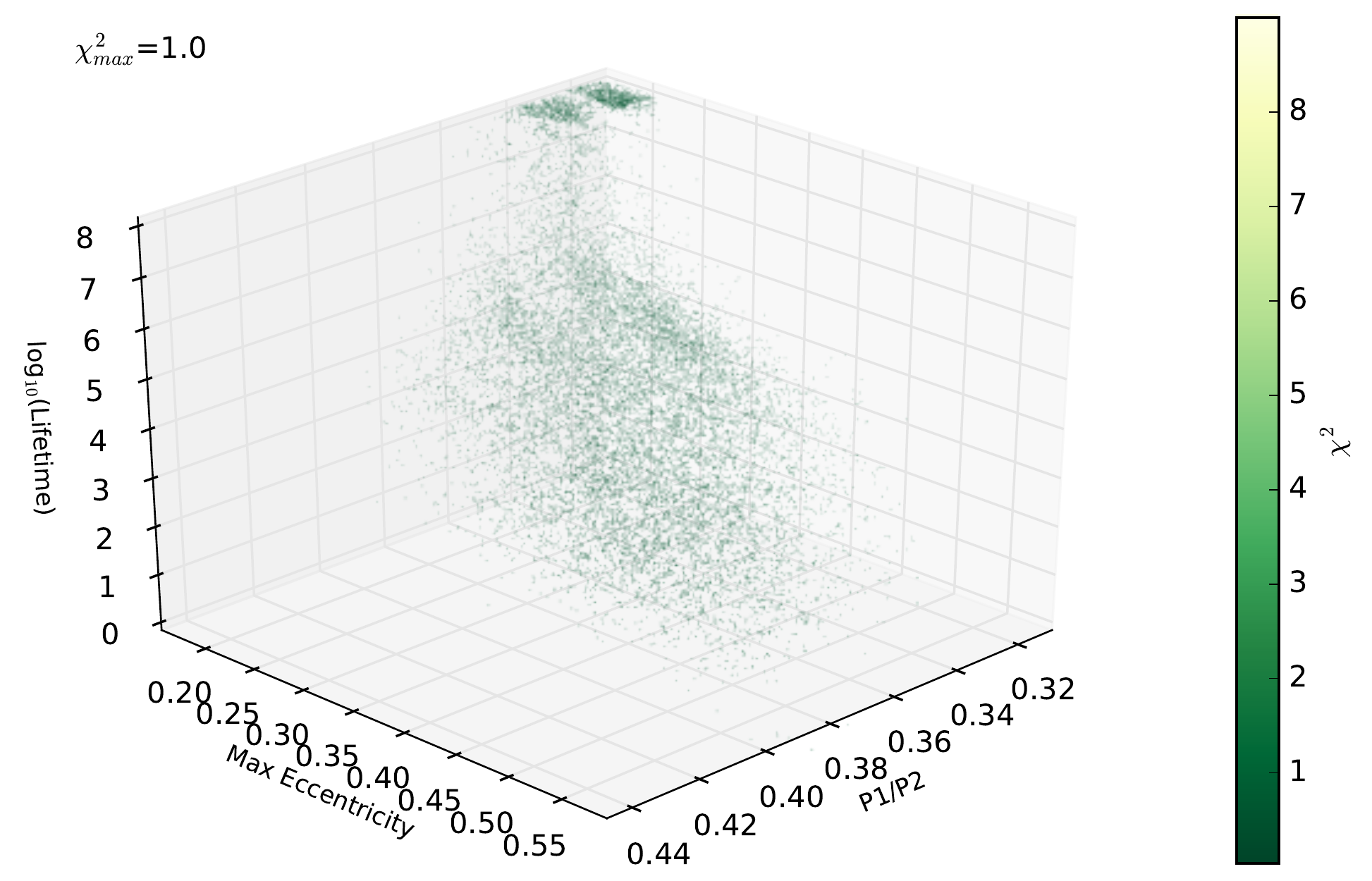}   \\
    \end{tabular}
\caption{The stability of the short-period solution for HD30177, as a 
function of the mean (left) and maximum (right) eccentricity of the two 
planets in the system. The colour axis shows the goodness of fit for 
each of the solutions tested, with the vertical axis showing the 
lifetime, and the y-axis the ratio of the two planetary orbital periods.  
The upper plots show the results for solutions within 3 $\sigma$ of the 
best fit, whilst the lower show only those simulations within 1 $\sigma$ 
of that solution.  We note that animated versions of the figures are 
available in the online edition of this work, which may help the reader 
to fully visualise the relationship between the stability and the 
various variables considered.}
\label{SP_eccentricity}
\end{figure}

\begin{figure}[SP_mass]
\centering
  \begin{tabular}{@{}cc@{}}
    \includegraphics[width=.46\textwidth]{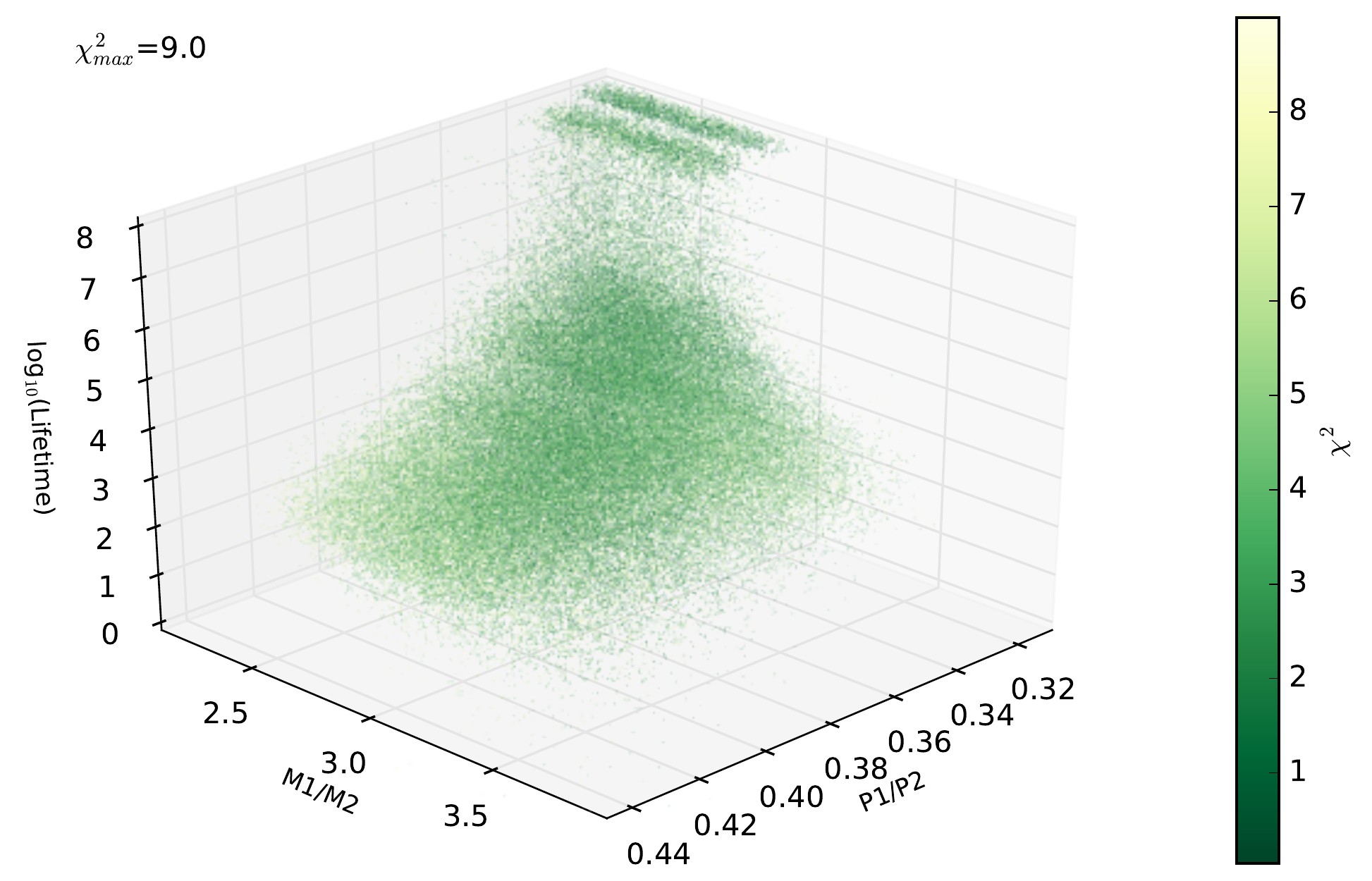} &
    \includegraphics[width=.46\textwidth]{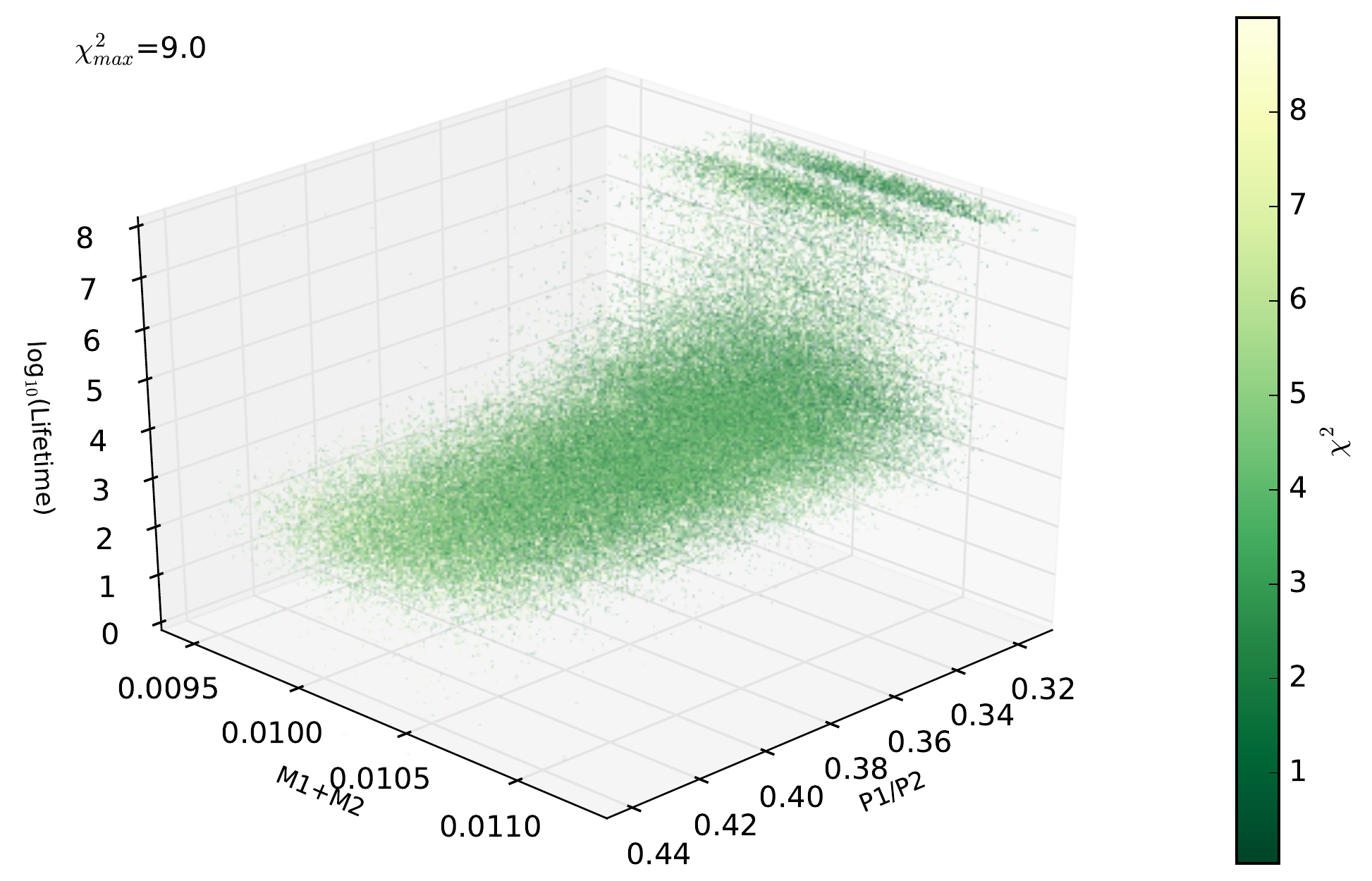} \\
    \includegraphics[width=.46\textwidth]{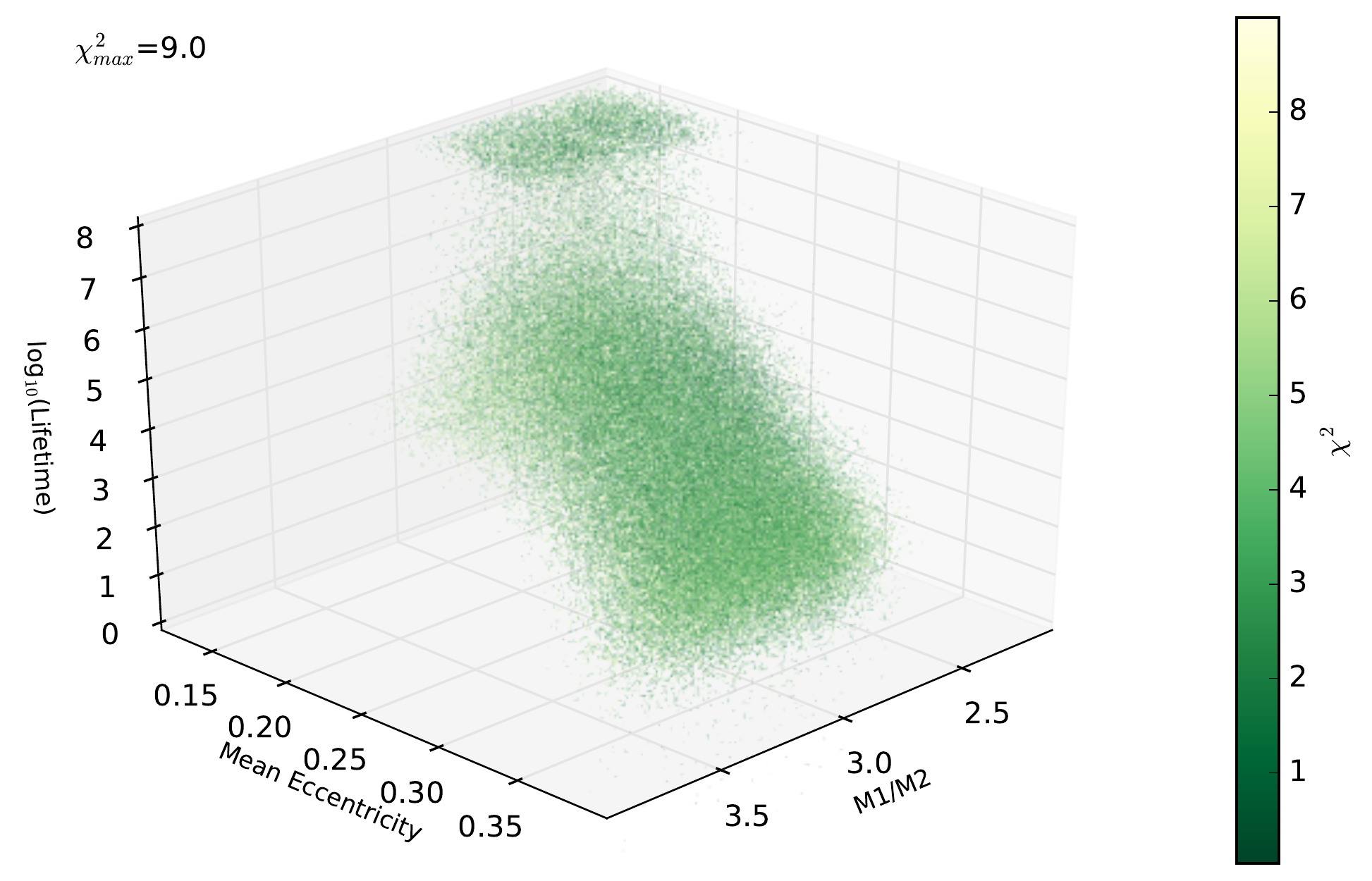} &
    \includegraphics[width=.46\textwidth]{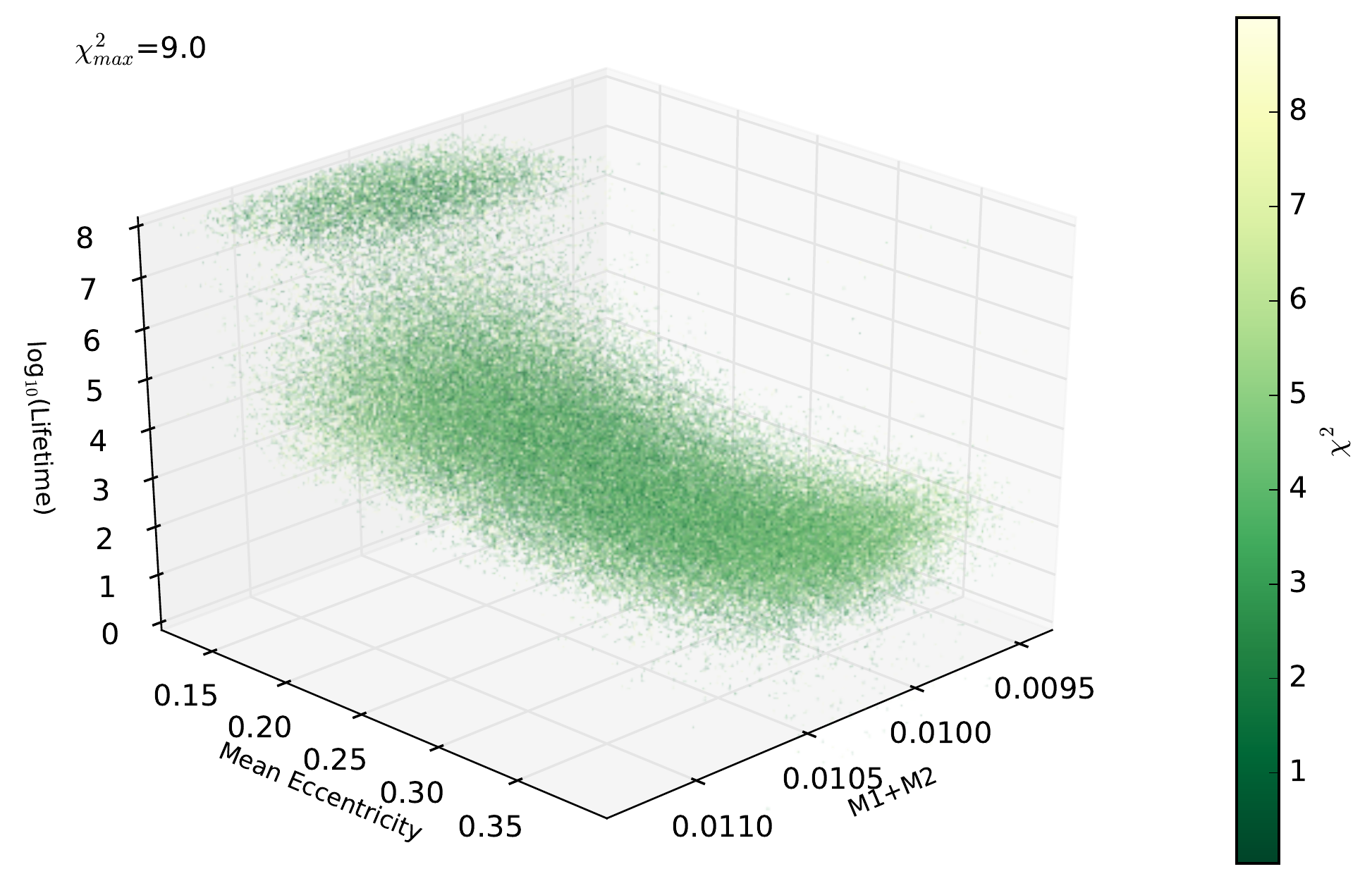} \\
    \includegraphics[width=.46\textwidth]{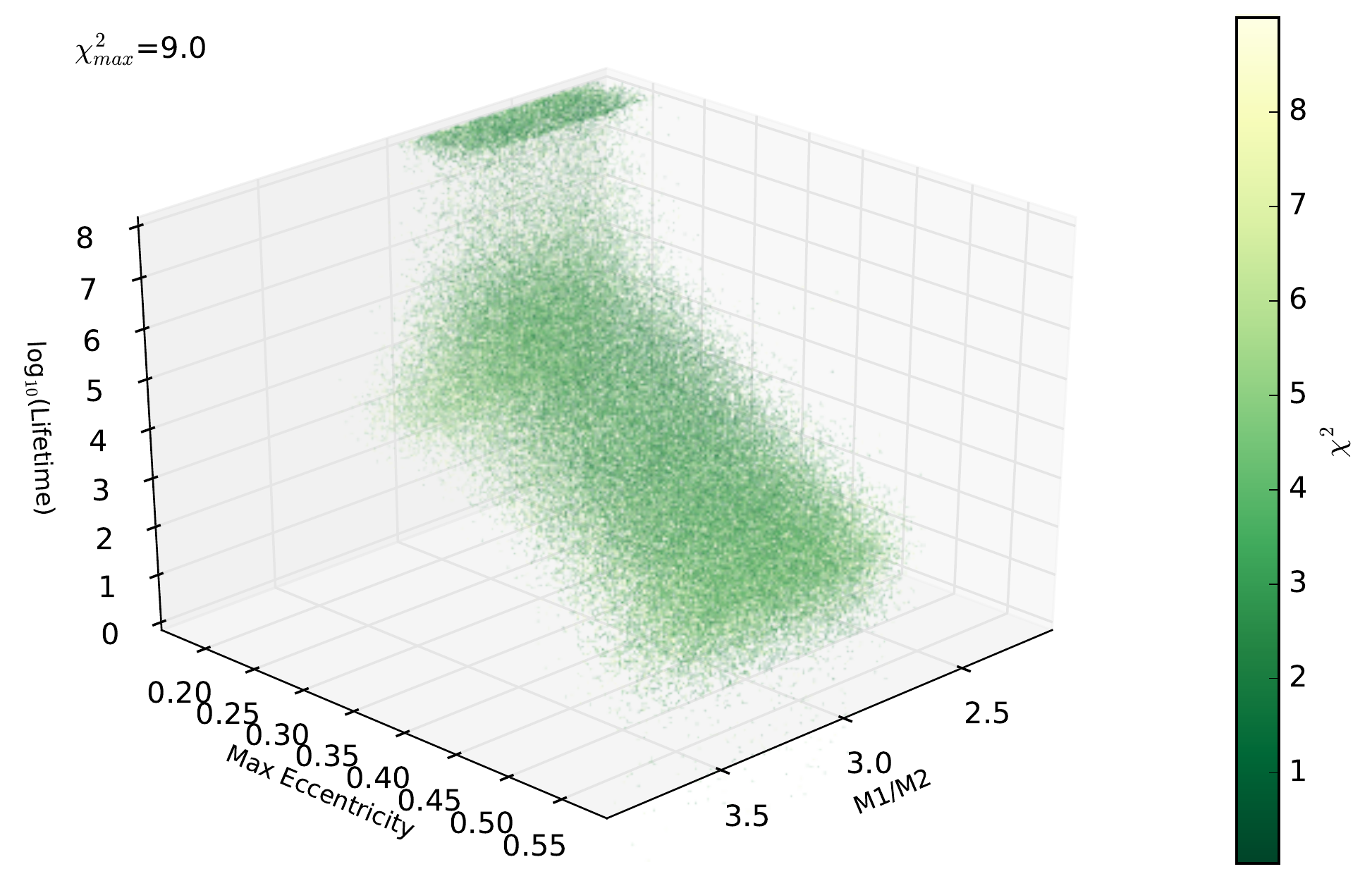} &
    \includegraphics[width=.46\textwidth]{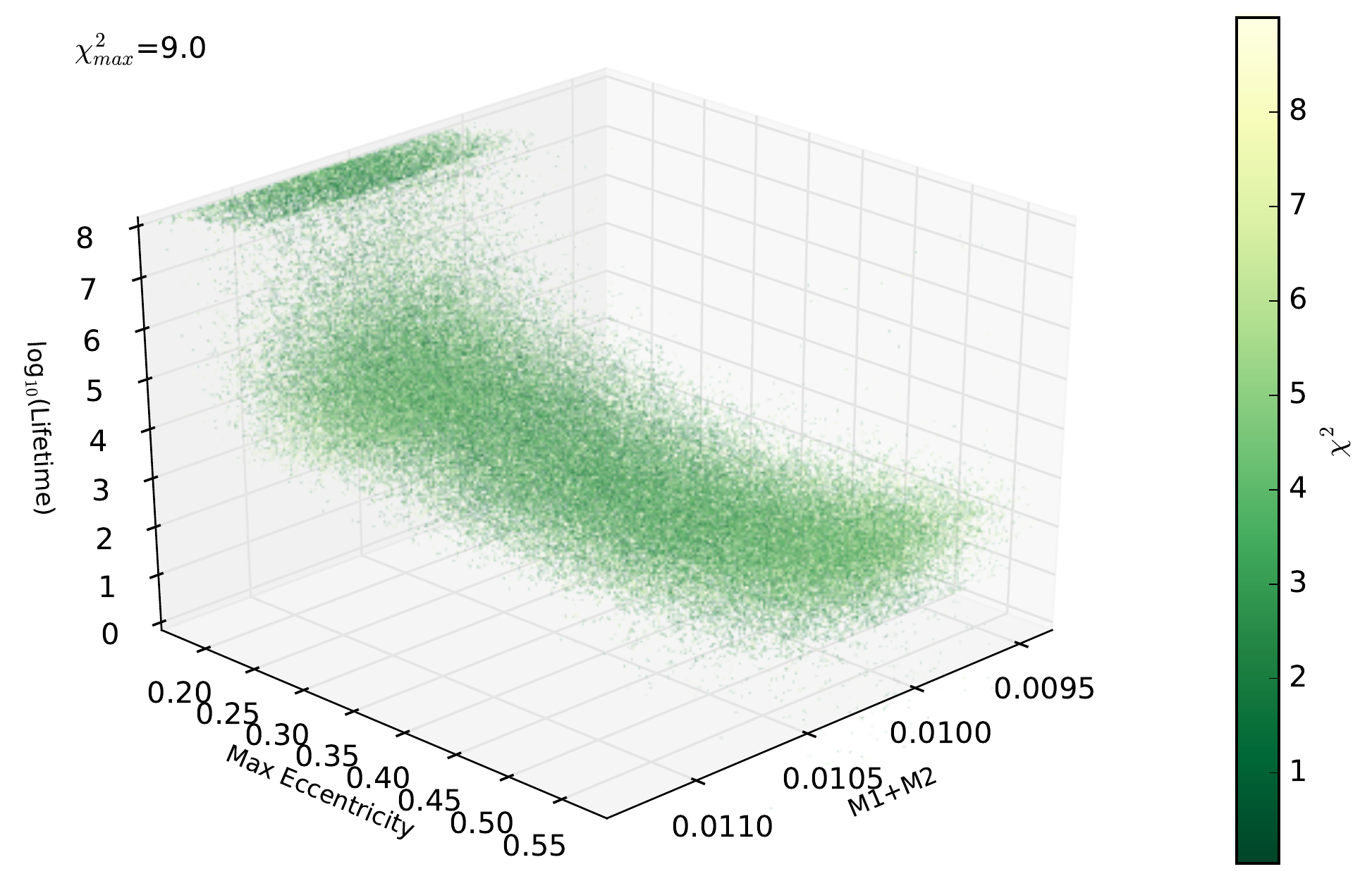} \\
    \end{tabular}
  \caption{\textit{Upper row:} The stability of the short-period 
solution for HD\,30177c, as a function of the mass ratio (left) and 
total mass (right) of the two planets in the system.  The color scale 
shows the goodness of fit for each of the solutions tested, with the 
vertical axis showing the lifetime, and the y-axis the ratio of the two 
planetary orbital periods.  Results for solutions within 3 $\sigma$ of 
the best fit are shown.  \textit{Middle row:} Same, but the x-axis now 
denotes the mean eccentricity of the planetary orbits.  \textit{Bottom 
row:} Same, but the x-axis now denotes the maximum eccentricity of the 
planetary orbits.  Animated versions of the figures are provided in the 
online version of the paper.}
\label{SP_mass}
\end{figure}

These results are strongly supported by our subsidiary integrations - 
the results of which are shown in 
Figures~\ref{SP_eccentricity}-\ref{SP_mass}. 
Figure~\ref{SP_eccentricity} shows the stability of the candidate 
HD\,30177 planetary systems as a function of the eccentricities of the 
two planets, their period ratio, and the goodness of the fit of the 
solution tested.  In Figure~\ref{SP_eccentricity}, the upper 
plots show all solutions within $3 \sigma$ of the best fit, whilst the 
lower show only those solutions within $1 \sigma$ of the best fit.  It 
is immediately apparent that truly stable solutions are limited to only 
a very narrow range of the plots - namely two narrow regions with low 
eccentricities, and widely separated orbits.  In fact, these solutions 
all lie at, or somewhat outside, the location of the 3:1 mean motion 
resonance between the two planets (P1/P2 $\sim 0.33$).  The inner of the 
two stable patches are those orbits that are resonant, whilst the 
outermost are those sufficiently separated to be exterior to that 
resonance.  Even at these stable separations, the system is only 
feasible for low-to-moderate planetary eccentricities - solutions that 
ascribe an eccentricity of $\sim 0.25$ or greater to either planet prove 
strongly unstable.

Figure~\ref{SP_mass} shows the influence of the mass of the two planets 
on the stability of the solution.  The resonant and extra-resonant 
stable regions are again clearly visible, and it is apparent that the 
masses of the two planets seem to have little influence on the stability 
of the solution.  A slight influence from the planetary mass can be seen 
in the middle row of Figure~\ref{SP_mass}, which shows that stable 
solutions with the lowest cumulative planet mass (i.e. $M_b + M_c$) have 
slightly higher mean eccentricities than those for larger cumulative 
masses.  This effect is only weak, and is the result of the least 
massive solutions veering away from lower eccentricities.  Given that 
the eccentricities of planetary orbits tend to be somewhat 
over-estimated when fitting radial velocity data \citep{otoole09}, this 
may be an indication that the lower-mass solutions are slightly less 
favourable than their higher mass counterparts.

Finally, the bottom row of Figure~\ref{SP_mass} shows the stability of 
the solution clouds as a function of the maximum eccentricity between 
the two planets (i.e. the value plotted on the y-axis is whichever is 
greater of $e_b$ and $e_c$).  This reinforces the result from 
Figure~\ref{SP_eccentricity} that solutions with either of the two 
planets moving on orbits with $e \geq 0.25$ are unstable regardless of 
their separation, or the mass of the planets involved.

Taken in concert, our results show that, whilst short-period solutions 
for the HD\,30177 system can prove dynamically stable, they require the 
two planets to either be trapped in mutual 3:1 mean motion resonance, or 
to be more widely spaced, and further require that neither planet move 
on an orbit with eccentricity greater than 0.25.

\begin{figure}[LP_eccentricity]
\centering
  \begin{tabular}{@{}cc@{}}
    \includegraphics[width=.46\textwidth]{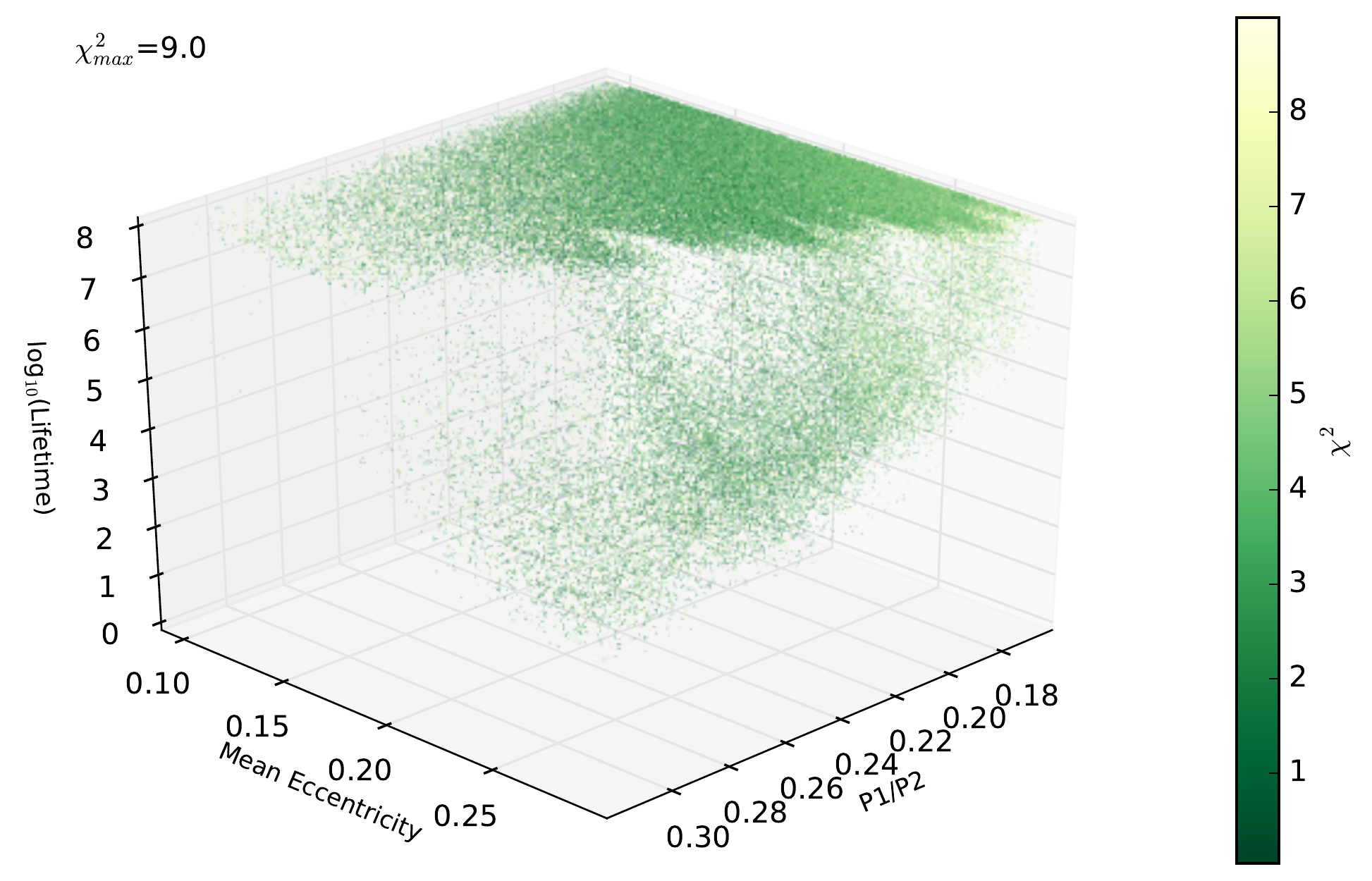} &
    \includegraphics[width=.46\textwidth]{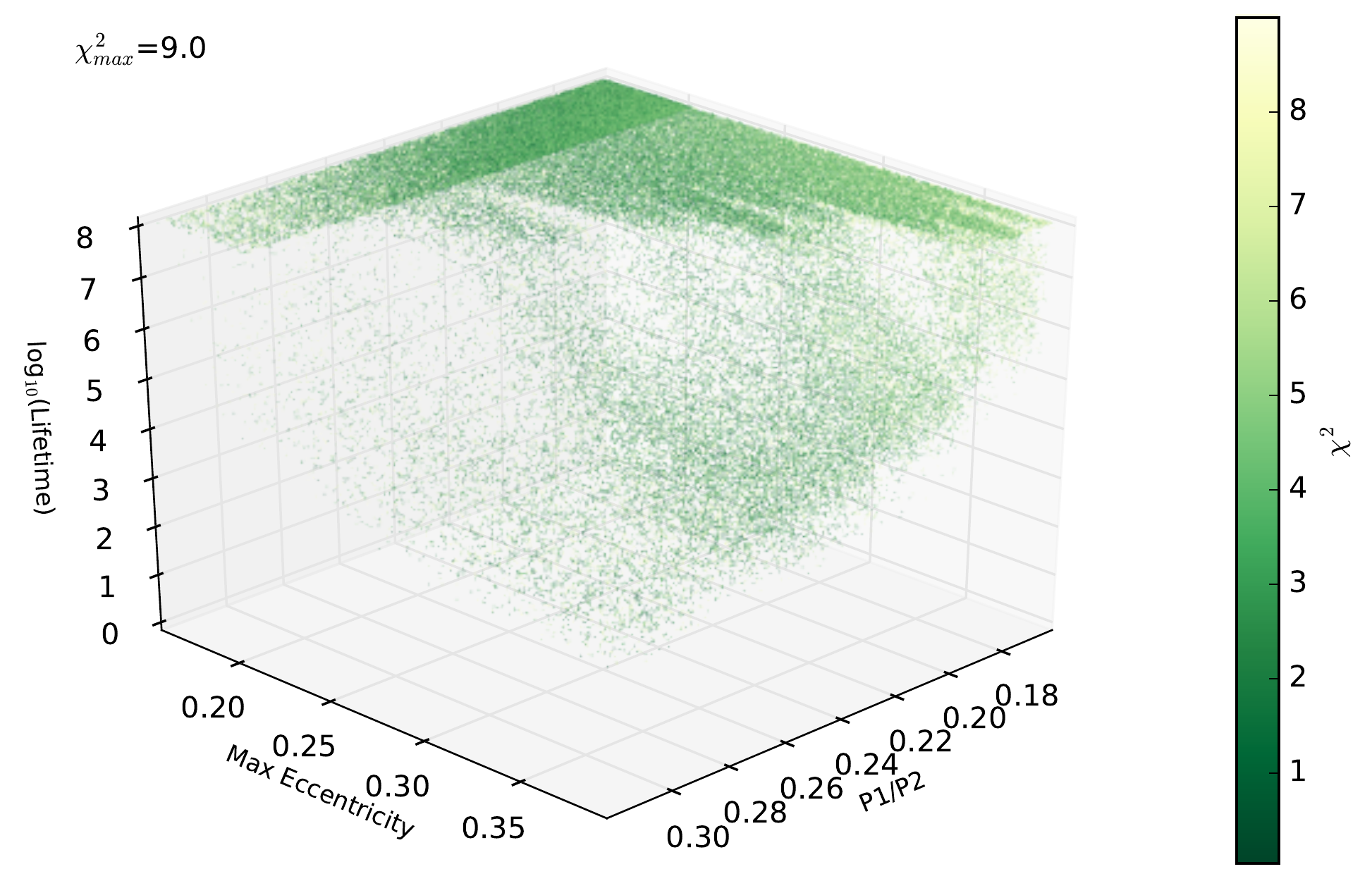} \\
    \end{tabular}
  \caption{The stability of the long-period solution for HD\,30177, as a 
function of the mean (left) and maximum (right) eccentricity of the two 
planets in the system.  The color scale and axes have the same meaning 
as in Figure~\ref{SP_eccentricity}.  The upper plots show the results 
for solutions within 3 $\sigma$ of the best fit, whilst the lower show 
only those simulations within 1 $\sigma$ of that solution.  Animated 
versions of the figures are provided in the online version of the 
paper.}
\label{LP_eccentricity}
\end{figure}

\begin{figure}[LP_mass]
\centering
  \begin{tabular}{@{}cc@{}}
    \includegraphics[width=.46\textwidth]{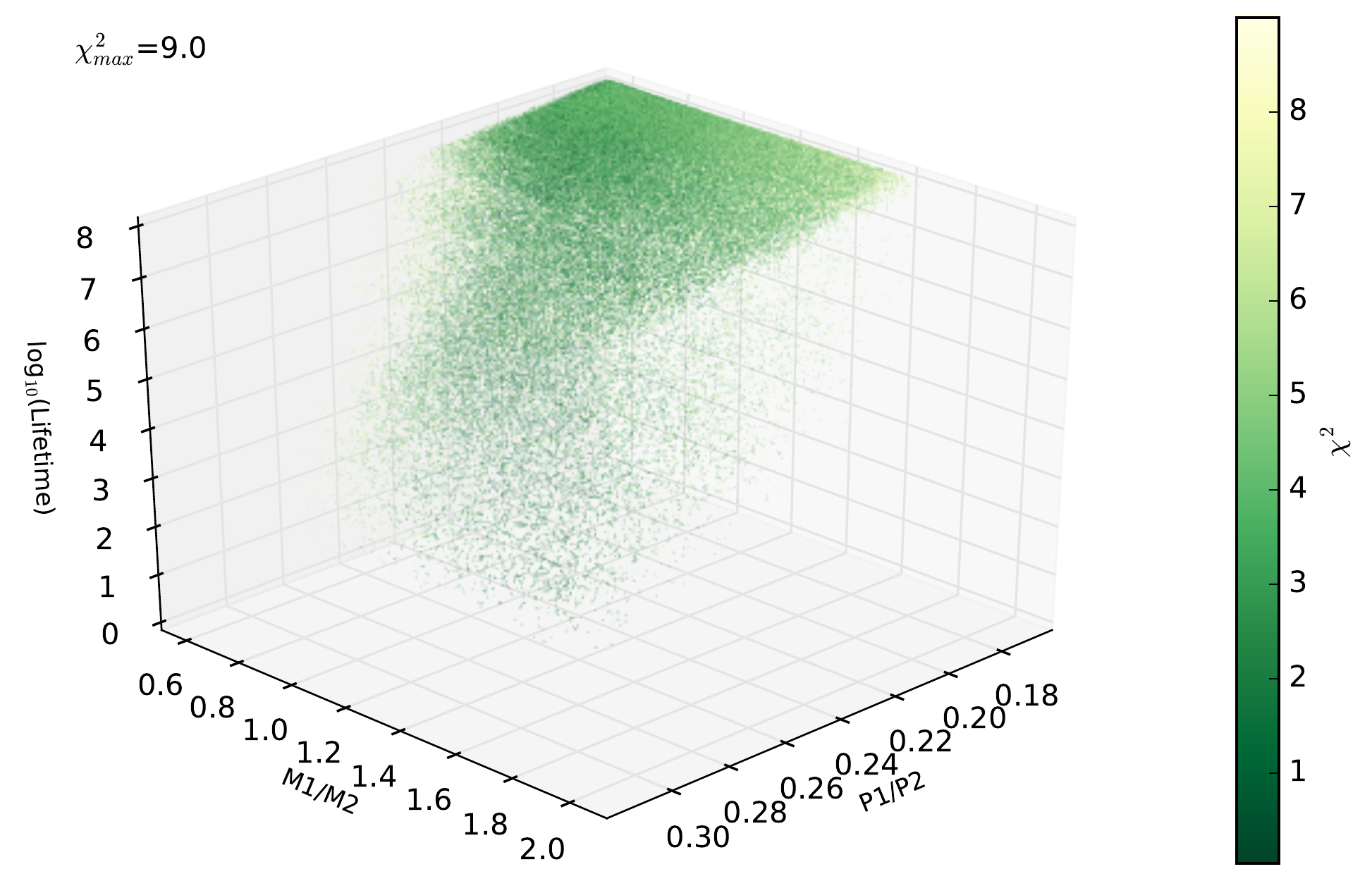} &
    \includegraphics[width=.46\textwidth]{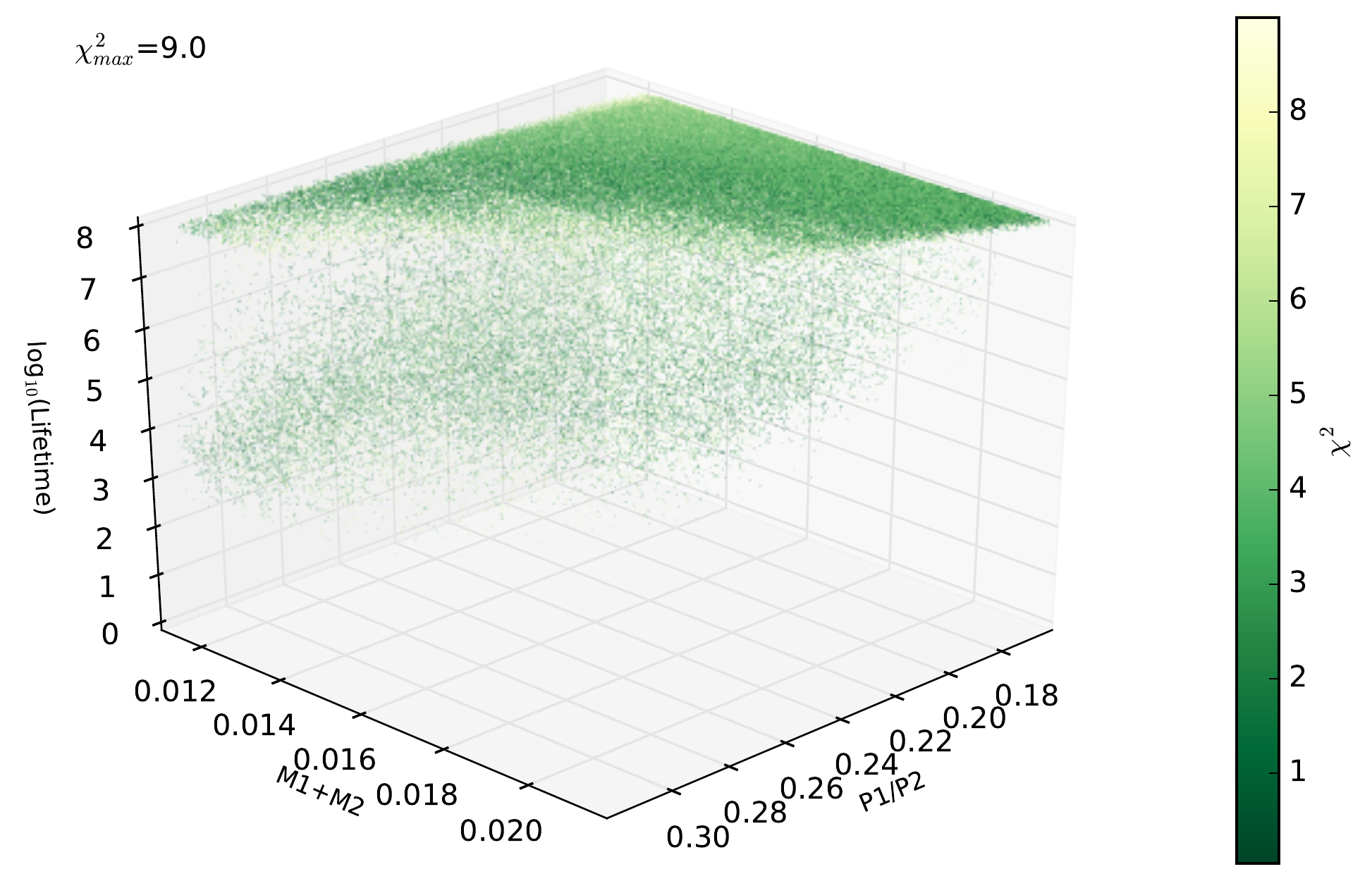} \\
    \includegraphics[width=.46\textwidth]{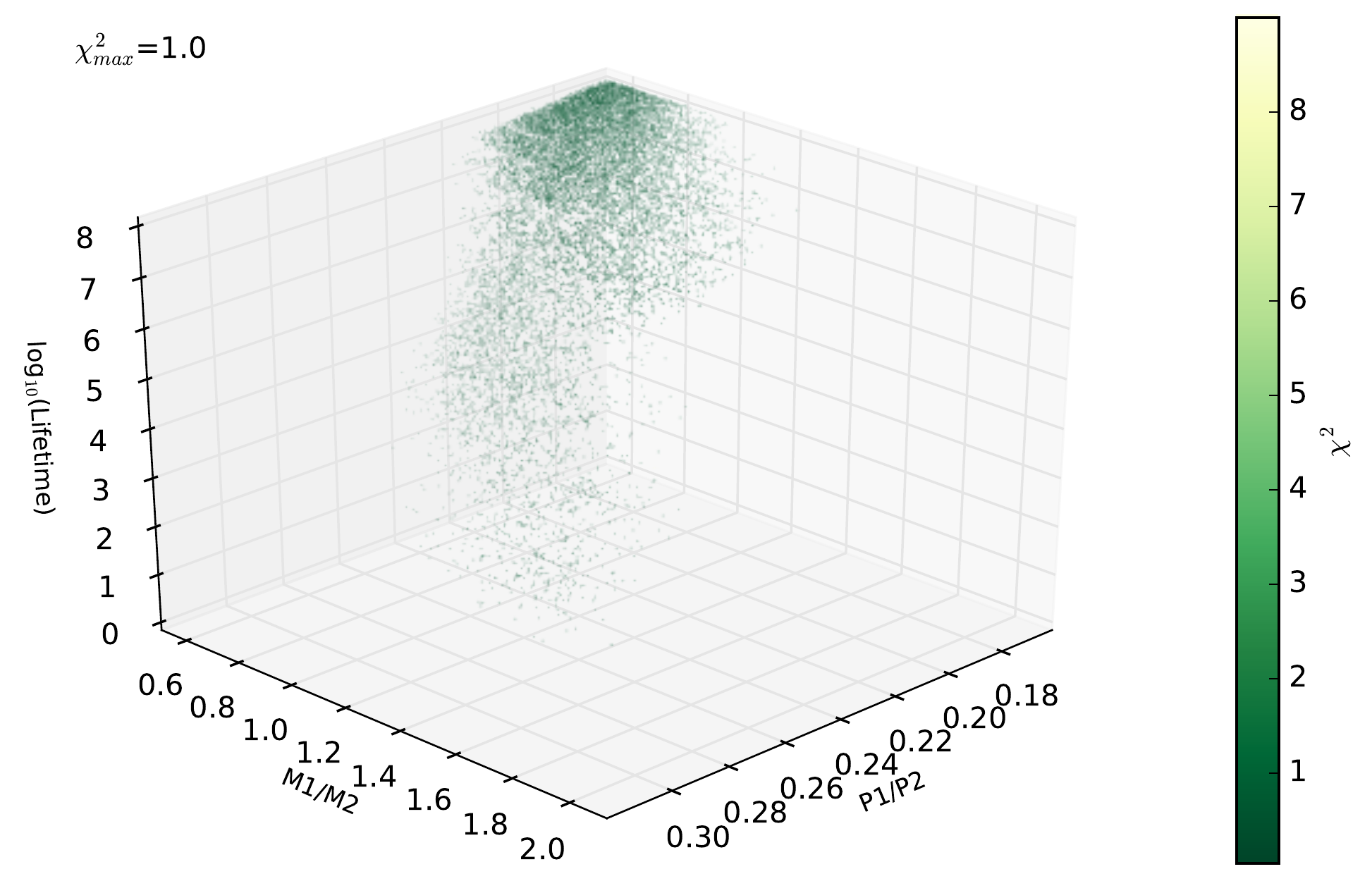} &
    \includegraphics[width=.46\textwidth]{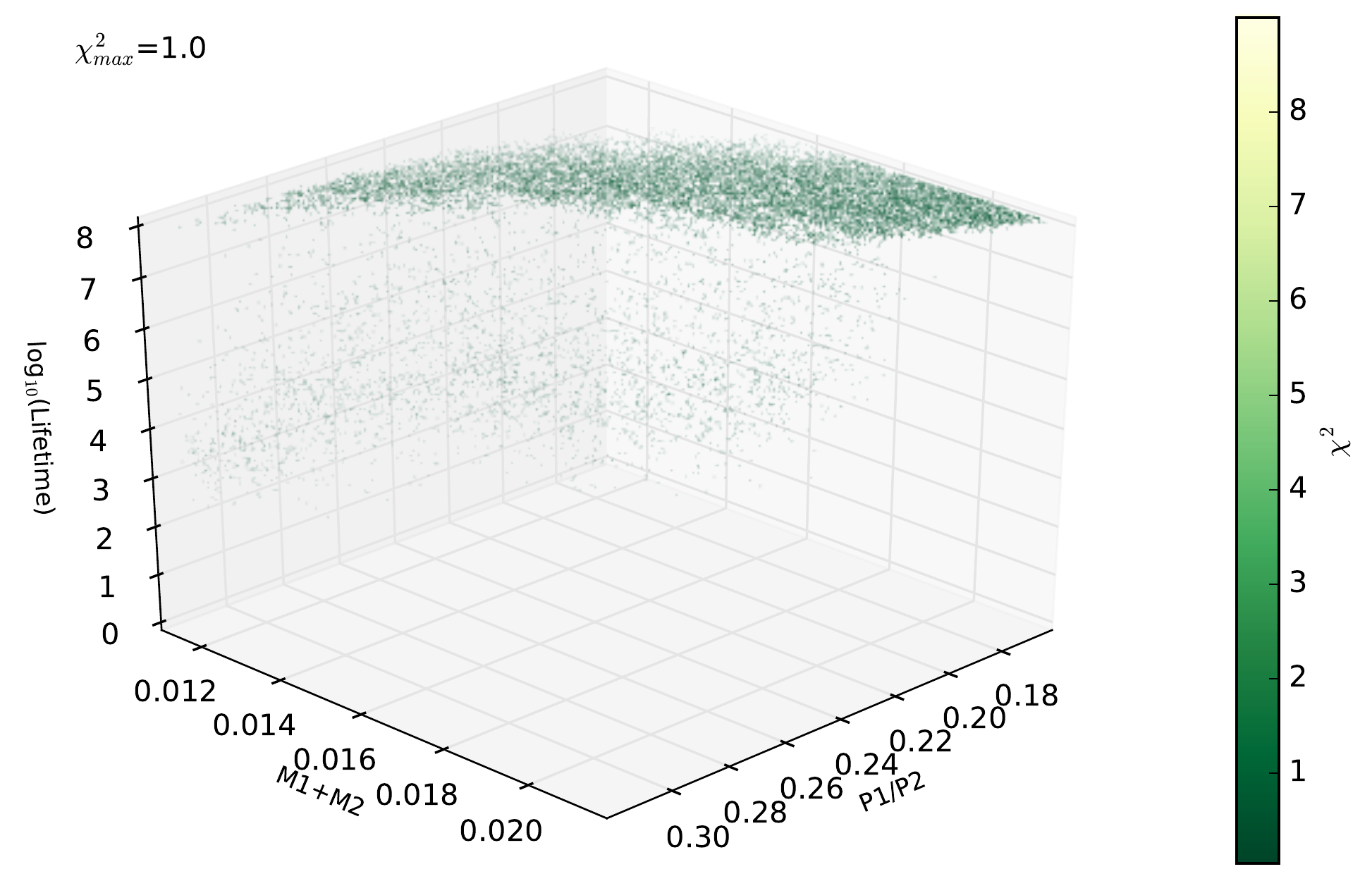}   \\
    \end{tabular}
  \caption{The stability of the long-period solution for HD\,30177, as a 
function of the mass ratio (left) and total mass (right) of the two 
planets in the system. As before, the colour axis shows the goodness of 
fit for each of the solutions tested, with the vertical axis showing the 
lifetime, and the y-axis the ratio of the two planetary orbital periods. 
Solutions within 3 $\sigma$ of the best fit are shown in the upper 
panels, and only those within 1$\sigma$ are shown in the lower panels.  
Animated versions of the figures are provided in the online version of 
the paper.}
\label{LP_mass}
\end{figure}

\begin{figure}[LP_mvse_mean]
\centering
  \begin{tabular}{@{}cc@{}}
    \includegraphics[width=.46\textwidth]{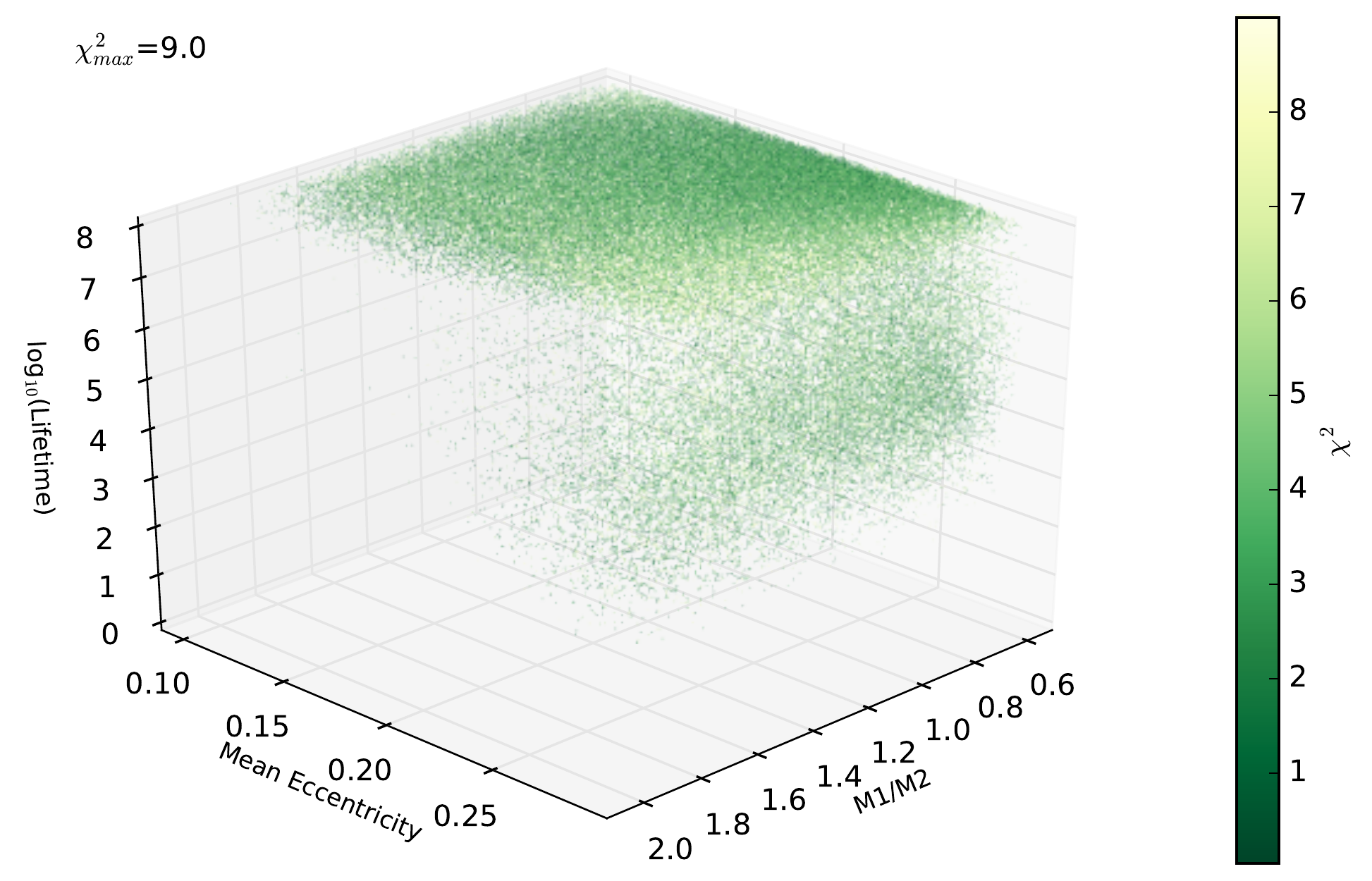} &
    \includegraphics[width=.46\textwidth]{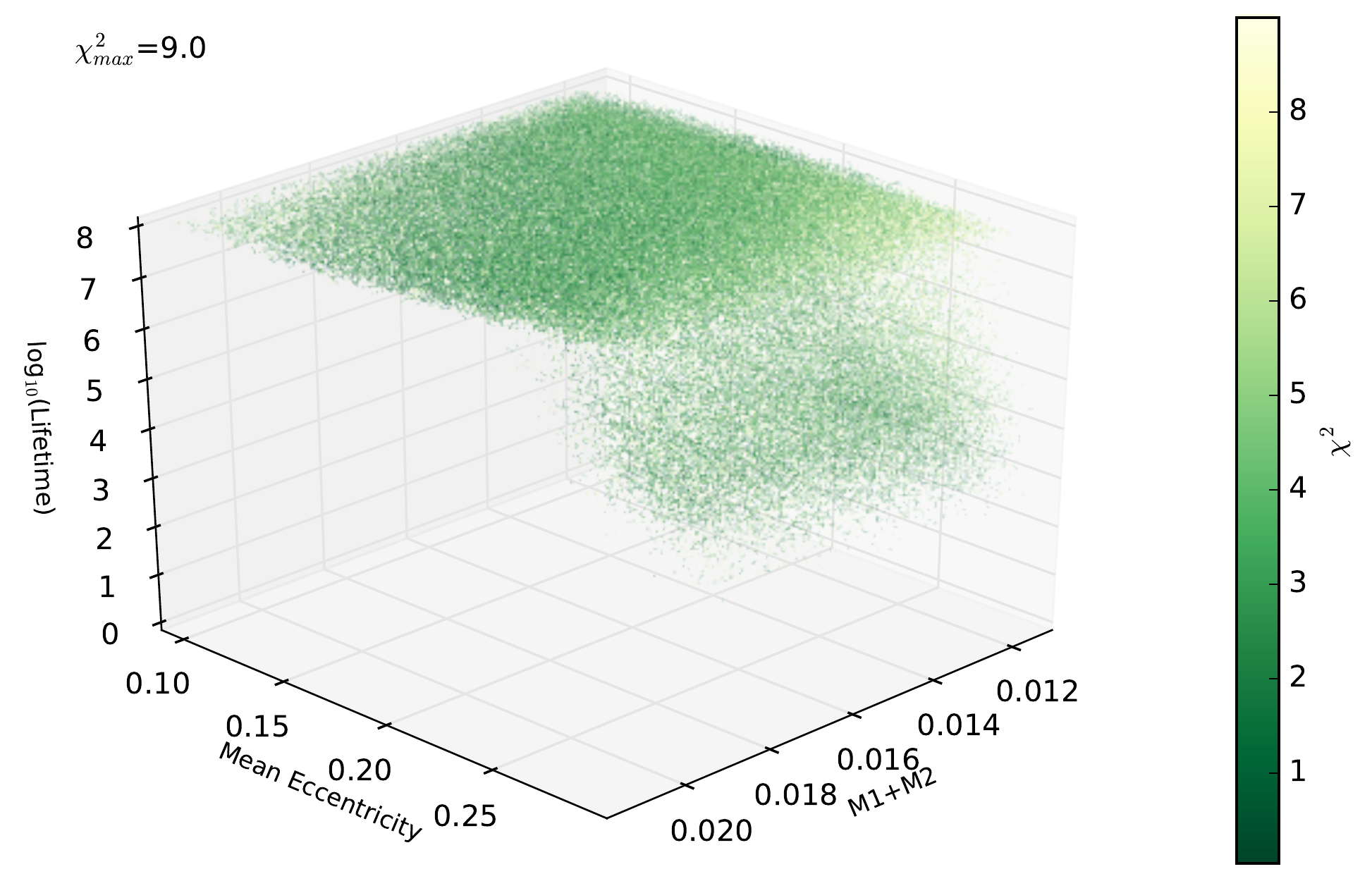} \\
    \end{tabular}
  \caption{The stability of the long-period solution for HD\,30177, 
again as a function of the mass ratio (left) and total mass (right) of 
the two planets in the system. Again, the colour axis shows the goodness 
of fit for each of the solutions tested, with the vertical axis showing 
the lifetime, and the x-axis the mean eccentricity of the planetary 
orbits.  Results for solutions within 3 $\sigma$ of the best fit are 
shown.  Animated versions of the figures are provided in the online 
version of the paper.}
\label{LP_mvse_mean}
\end{figure}

\begin{figure}[LP_mvse_max]
\centering
  \begin{tabular}{@{}cc@{}}
    \includegraphics[width=.46\textwidth]{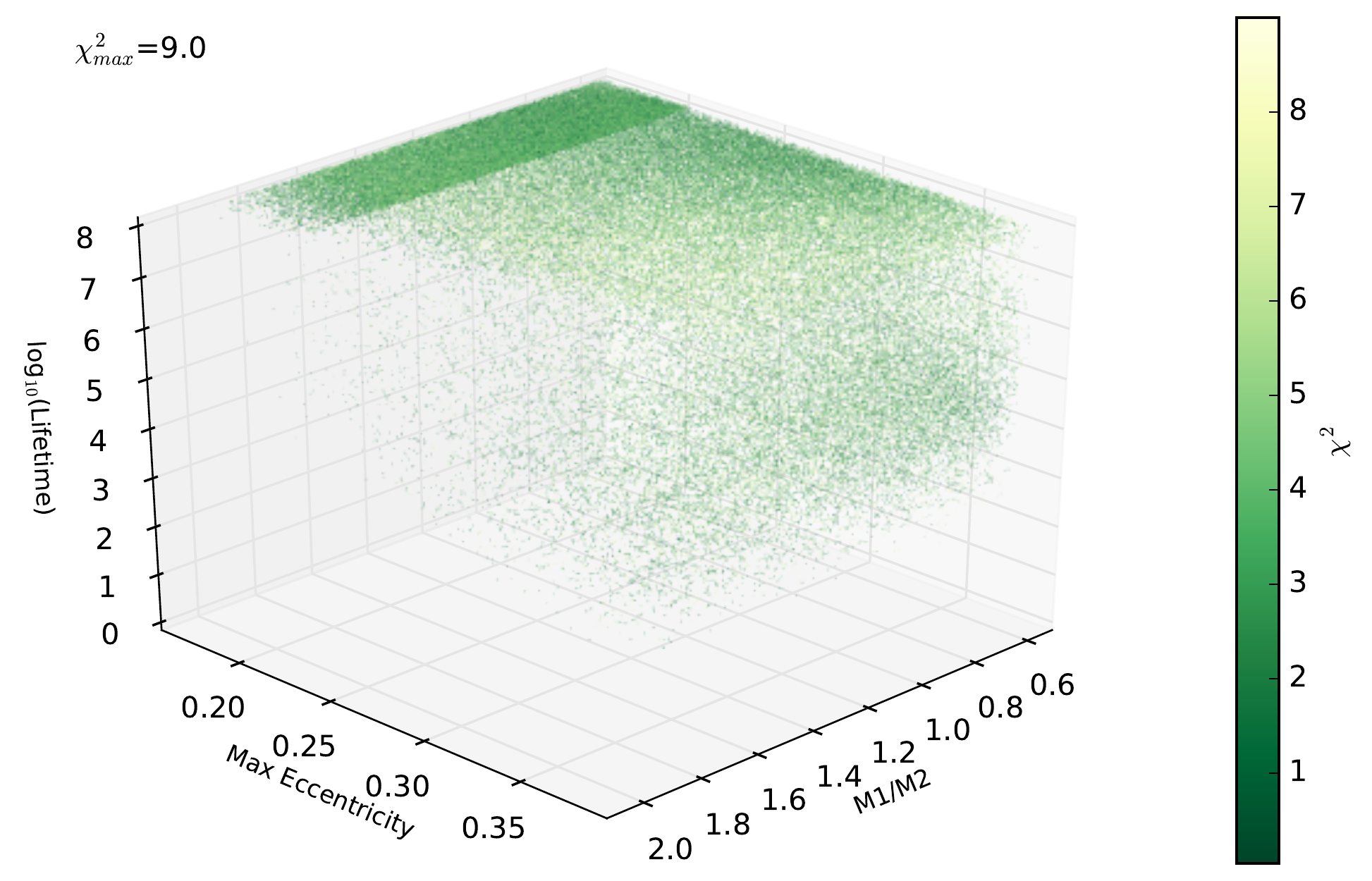} &
    \includegraphics[width=.46\textwidth]{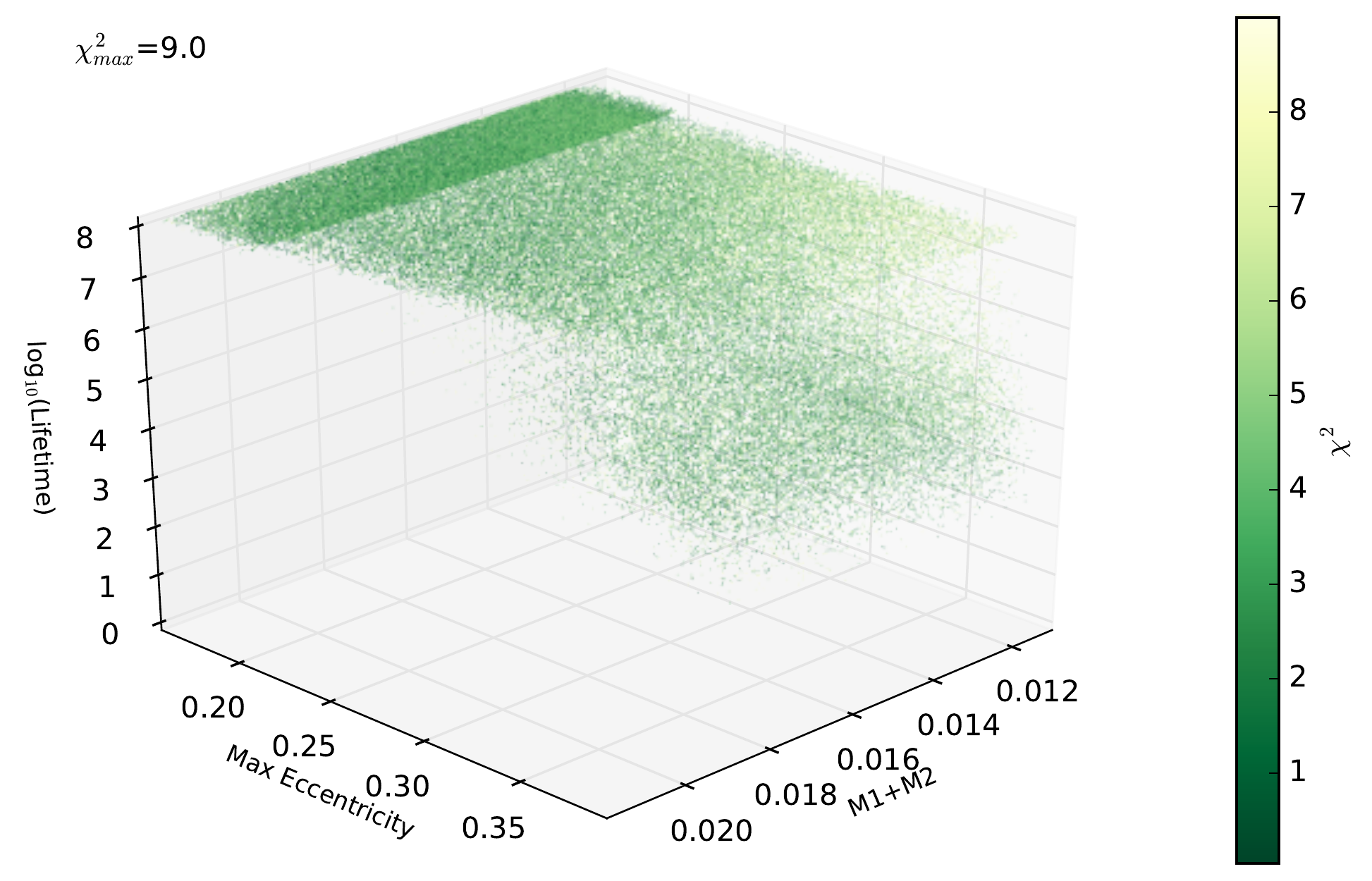} \\
    \includegraphics[width=.46\textwidth]{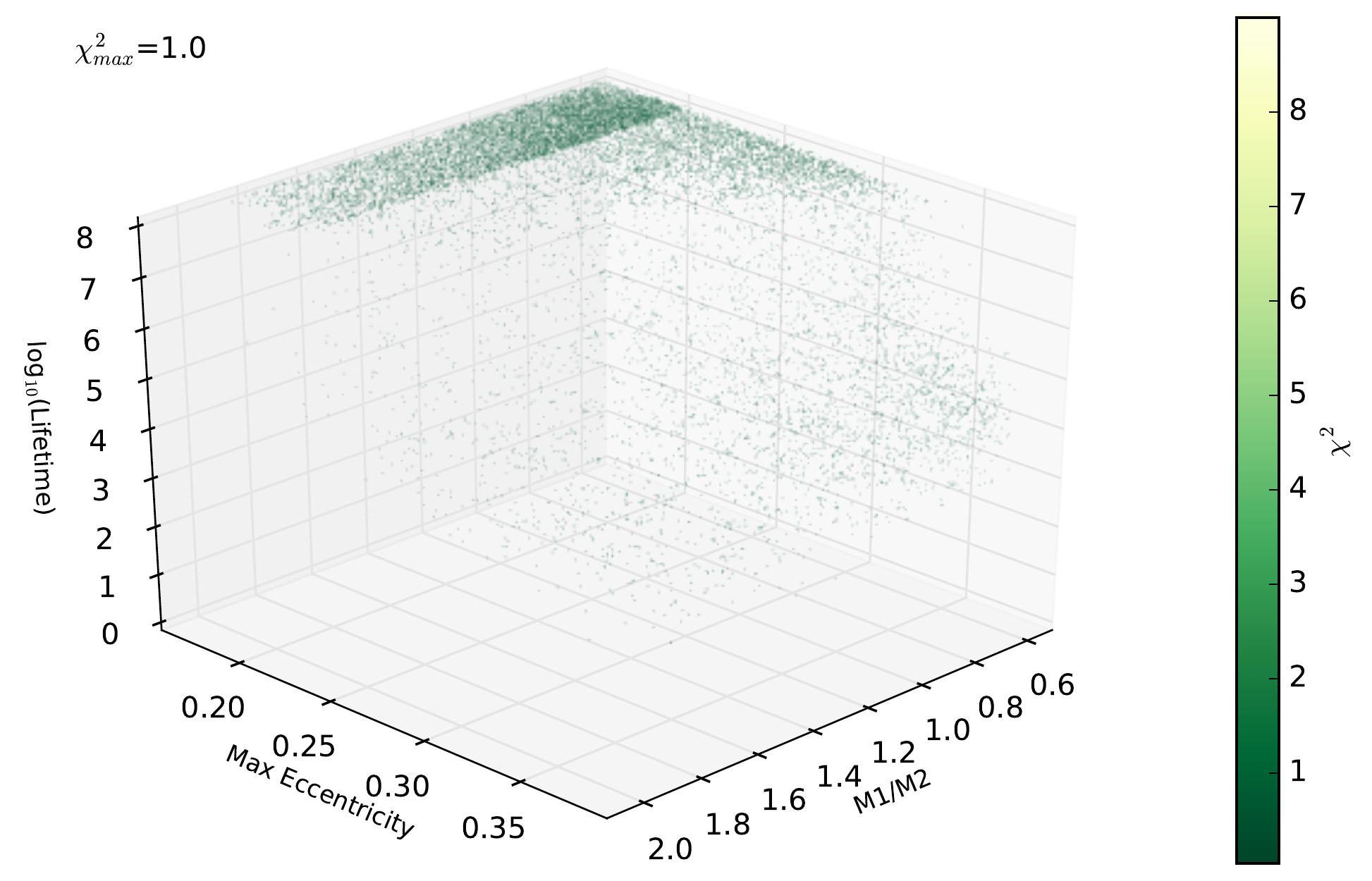} &
    \includegraphics[width=.46\textwidth]{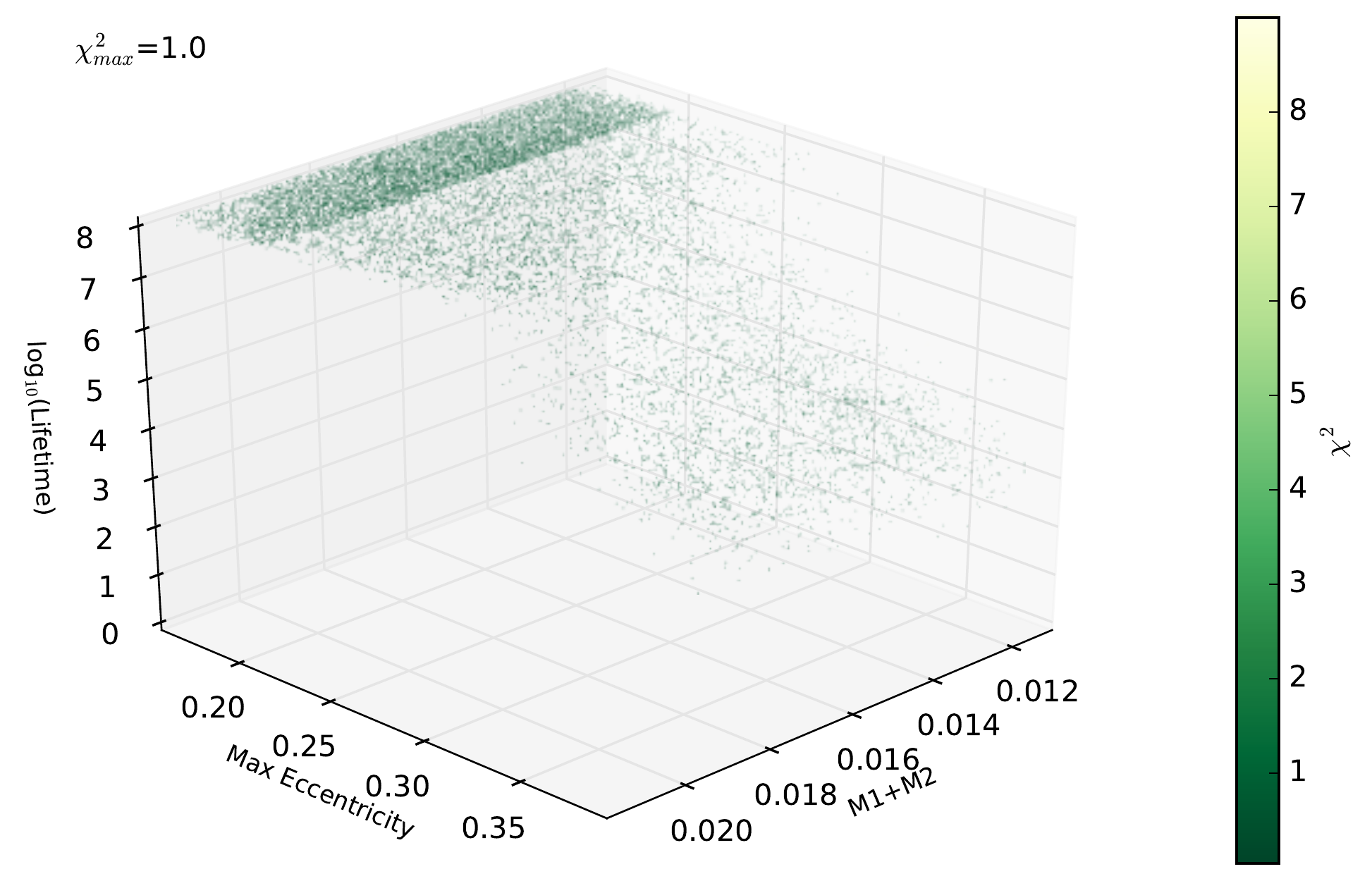}   \\
    \end{tabular}
  \caption{The stability of the long-period solution for HD\,30177, 
again as a function of the mass ratio (left) and total mass (right) of 
the two planets in the system.  The colour axis shows the goodness of 
fit for each of the solutions tested, with the vertical axis showing the 
lifetime, and the x-axis the maximum eccentricity of the planetary 
orbits.  The upper plots show the results for solutions within 3 
$\sigma$ of the best fit, whilst the lower show only those simulations 
within 1 $\sigma$ of that solution.  Animated versions of the figures 
are provided in the online version of the paper.}
\label{LP_mvse_max}
\end{figure}

But what of our alternative, longer-period solution for the planetary 
system?  Figure~\ref{LongPeriodOrbit} shows the dynamical context of 
that solution.  Unlike the short period solution, the two planets are 
now sufficiently widely separated that the great majority of orbits 
around the best-fit solution are now dynamically stable for the full 100 
Myr of our simulations.  At the very inner edge of the plot, the cliff 
of instability exterior to the planet's mutual 3:1 mean-motion resonance 
can again be seen, as can hints of the moderate stability afforded by 
the periastron of HD\,30177 c falling at the semi-major axis of 
HD\,30177 b (top left of the plot).  Purely on the basis of this plot, 
the longer-period solution seems markedly more dynamically feasible, a 
result once again borne out by the plots of our subsidiary simulations 
of the long-period solution 
(Figures~\ref{LP_eccentricity}-\ref{LP_mvse_max}).

Figure~\ref{LP_eccentricity} reveals many of the same features as 
Figure~\ref{LongPeriodOrbit} - a significant proportion of the solutions 
are dynamically stable - particularly those within $1 \sigma$ of the 
best fit (lower panels).  The greater the orbital separation of the two 
planets, the greater can be their orbital eccentricities before 
destabilising the system.  In addition, however, the destabilising 
influence of distant mean-motion resonances is revealed in these plots, 
as the notches of instability carved into the distribution at specific 
period ratios.  Aside from these few unstable regions, however, the 
great majority of solutions within $1 \sigma$ of the best fit are 
stable.

Figure~\ref{LP_mass} shows that the mass ratio of the planets (left hand 
plots) has little or no influence on their stability. Interestingly, 
though, the lower-right hand panel reveals an apparent relationship in 
the fitting between the cumulative mass of the planets and their mutual 
separation.  The more widely separated the two planets (and hence the 
more distant is HD 30177c), the greater their cumulative mass. This is 
not at all surprising: the more distant HD 30177c is, the greater its 
mass would have to be to achieve a radial velocity signal of a given 
amplitude.  This feature is therefore entirely expected, but 
nevertheless serves to nicely illustrate the relationship between 
different parameters in the radial velocity fitting process.

Figure~\ref{LP_mvse_mean} again reveals that the more eccentric the 
orbits of the planets, the more likely they are to prove unstable - 
although once again, the great majority of the sampled phase-space 
proves dynamically stable.  More interesting, however, are the results 
shown in Figure~\ref{LP_mvse_max}.  The left-hand panels of that plot, 
which show the stability of the solutions as a function of the maximum 
eccentricity between the two panels (y-axis) and the mass ratio of the 
two planets (x-axis) suggest that, the closer the two planetary masses 
are to parity, the more likely eccentric orbits are to be stable.  By 
contrast, the lower-right hand plot suggests that the greater the sum of 
the planetary masses, the more likely solutions with high eccentricities 
are to be stable.  Taken in concert, these results are once again a 
reflection of the relationship between cumulative mass and orbital 
separation.  That is, the greater the orbital separation of the two 
planets, the greater their cumulative mass, and the closer to parity 
their masses become (since our fits suggest that HD\,30177c is the less 
massive of the two).  At the same time, we saw from 
Figure~\ref{LP_eccentricity} that, the greater the separation of the two 
planets, the more stable are those orbital solutions at higher 
eccentricity.  So once again, we are looking at the same thing - these 
two apparent trends are the result of the requirement that a more 
distant HD\,30177c must be more massive in order to generate the 
observed radial velocity amplitude.

\section{Conclusions}

We present the results of new radial velocity observations of HD\,30177, 
which reveal for the first time the presence of a second, long-period 
planet in that system.  Two possible orbital solutions for the planetary 
system are presented - one with a shorter-period orbit for HD\,30177 c, 
and one with the two planets more widely spaced, and HD\,30177 c on a 
longer period orbit.  The two solutions are virtually indistinguishable 
from one another in terms of the quality of fit that they provide to the 
data.  However, the short-period solution placed the two planets on 
orbits sufficiently compact that they lie closer than their mutual 3:1 
mean-motion resonance.

Although several highly compact multi-planet systems have been 
discovered in recent years, it has become apparent that such compact 
systems rely on dynamical stability conferred by mutual resonant 
planetary orbits.  As such, it seemed prudent to build on our earlier 
work, and carry out detailed $n$-body simulations of the two potential 
solutions for the HD\,30177 system, to see whether it was possible to 
rule either out on dynamical stability grounds.

Our results reveal that, although some stable solutions can be found for 
the short-period variant of the HD\,30177 system, those solutions 
require orbital eccentricities for the planets that are typically 
smaller than given by the best fit solution, and require HD\,30177 c to 
be somewhat more distant than the best fit.  In other words, the require 
relatively low eccentricity orbits for that planet exterior to its 
mutual 3:1 mean-motion resonance with HD\,30177 b.  By contrast, the 
great majority of the longer-period solutions tested proved dynamically 
stable - and across a much greater range of potential semi-major axes 
and orbital eccentricities.

As a result, we consider that the most likely solution for the orbit of 
HD\,30177c is the longer period option: an m~sin~$i=7.6\pm$3.1\,\Mjup\ 
planet with $a = 9.89\pm1.04$~au, $e = 0.22\pm0.14$, and an orbital 
period of $P = 11613\pm1837$ days.  We note that for 
inclinations $i\ltsimeq$30\degrees, the two orbiting bodies in the 
HD\,30177 system fall into the brown dwarf regime.  With an orbital 
separation of $a\sim$10\,au, one can consider HD\,30177c to be one of 
the first members of an emerging class of ``Saturn analogs,'' referring 
to planets with orbital separations similar to Saturn.  Just as 
long-term radial velocity surveys have begun to characterize ``Jupiter 
analogs'' \citep{jupiters, rowan16, newjupiters}, the continuation of 
legacy surveys such as the Anglo-Australian Planet Search will enable us 
to probe the population of planets in Saturn-like orbits in the coming 
decade.

\acknowledgements

JH is supported by USQ's Strategic Research Fund: the STARWINDS project.  
CGT is supported by Australian Research Council grants DP0774000 and 
DP130102695.  This research has made use of NASA's Astrophysics Data 
System (ADS), and the SIMBAD database, operated at CDS, Strasbourg, 
France.  This research has also made use of the Exoplanet Orbit Database 
and the Exoplanet Data Explorer at exoplanets.org \citep{wright11, 
han14}.


\begin{deluxetable}{lll}
\tabletypesize{\scriptsize}
\tablecolumns{3}
\tablewidth{0pt}
\tablecaption{Stellar Parameters for HD 30177}
\tablehead{
\colhead{Parameter} & \colhead{Value} &  \colhead{Reference}
 }
\startdata
\label{stellarparams}
Spec.~Type & G8 V  & \citet{houk75} \\
Distance (pc) & 54.7$\pm$2.3 & \citet{vl07} \\
Mass (\Msun) & 0.951$^{+0.093}_{-0.053}$  & \citet{takeda07} \\
    & 1.05$\pm$0.08 & \citet{santos13} \\
    & 0.988$\pm$0.033 & \citet{sousa11} \\ 
V sin $i$ (\kms) & 2.96$\pm$0.50 & \citet{butler06} \\
$[Fe/H]$ & +0.33$\pm$0.05 & \citet{franchini14} \\
    & 0.37$\pm$0.06 & \citet{adi12} \\
    & 0.39$\pm$0.05 & \citet{ghezzi10} \\
    & 0.394$\pm$0.030 & \citet{butler06} \\
$T_{eff}$ (K) & 5580$\pm$12 & \citet{franchini14} \\
    & 5601$\pm$73 & \citet{adi12} \\
    & 5595$\pm$50 & \citet{ghezzi10} \\
    & 5607$\pm$44 & \citet{butler06} \\
log $g$ & 4.41$\pm$0.12 & \citet{franchini14} \\
    & 4.34$\pm$0.05 & \citet{sousa11} \\
    & 4.15$\pm$0.13 & \citet{ghezzi10} \\
    & 4.31$\pm$0.06 & \citet{butler06} \\
Age (Gyr) & 11.6$^{+1.8}_{-2.2}$ & \citet{takeda07} \\
\enddata
\end{deluxetable}

\begin{deluxetable}{lrr}
\tabletypesize{\scriptsize}
\tablecolumns{3}
\tablewidth{0pt}
\tablecaption{AAT Radial Velocities for HD 30177}
\label{AATvels}
\tablehead{
\colhead{BJD-2400000} & \colhead{Velocity (\ms)} & \colhead{Uncertainty
(\ms)}}
\startdata
51118.09737  &     227.2  &    4.5  \\
51119.19240  &     188.6  &    6.9  \\
51121.15141  &     210.7  &    6.1  \\
51157.10219  &     223.5  &    4.5  \\
51211.98344  &     234.6  &    5.0  \\
51212.96597  &     235.8  &    4.2  \\
51213.99955  &     245.4  &    4.0  \\
51214.95065  &     237.3  &    3.6  \\
51525.99718  &     177.1  &    3.4  \\
51630.91556  &     144.9  &    4.6  \\
51768.32960  &      73.4  &    4.2  \\
51921.10749  &      14.8  &    4.6  \\
52127.32049  &      -9.2  &    8.5  \\
52188.25324  &     -41.3  &    3.6  \\
52358.91806  &     -45.6  &    3.8  \\
52598.18750  &     -49.8  &    2.0  \\
52655.02431  &     -57.6  &    4.4  \\
52747.84861  &     -49.0  &    2.2  \\
52945.18132  &     -12.7  &    2.6  \\
53044.03464  &       8.2  &    3.6  \\
53282.26188  &      58.2  &    2.8  \\
53400.99440  &      91.4  &    2.5  \\
54010.25007  &     137.4  &    1.8  \\
54038.21420  &     126.4  &    3.4  \\
54549.93698  &     -47.3  &    2.2  \\
54751.25707  &     -83.8  &    3.8  \\
54776.17955  &     -79.6  &    2.2  \\
54900.95132  &     -78.0  &    3.4  \\
55109.18072  &     -77.0  &    3.2  \\
55457.26529  &     -32.2  &    3.9  \\
55461.28586  &     -25.3  &    4.3  \\
55519.17942  &      -2.1  &    3.3  \\
55845.21616  &      82.2  &    4.7  \\
55899.10987  &      79.0  &    3.2  \\
56555.28257  &     149.0  &    4.1  \\
56556.25219  &     152.0  &    3.6  \\
56746.90702  &      97.7  &    5.1  \\
56766.86295  &      66.2  &    4.0  \\
56935.25257  &      10.2  &    4.0  \\
56970.23271  &       5.6  &    3.0  \\
57052.02821  &      -2.2  &    3.0  \\
57094.90039  &     -28.0  &    4.6  \\
57349.14648  &     -34.5  &    3.1  \\
\enddata
\end{deluxetable}

\begin{deluxetable}{lrr}
\tabletypesize{\scriptsize}
\tablecolumns{3}
\tablewidth{0pt}
\tablecaption{HARPS-TERRA Radial Velocities for HD 30177}
\label{harpsvels}
\tablehead{
\colhead{BJD-2400000} & \colhead{Velocity (\ms)} & \colhead{Uncertainty
(\ms)}}
\startdata
52947.76453  &     -70.7  &    1.6  \\
53273.88347  &       0.0  &    1.9  \\
53274.88548  &       4.6  &    1.9  \\
53288.84830  &       4.6  &    1.5  \\
53367.68146  &      21.6  &    1.8  \\
53410.60057  &      32.3  &    1.5  \\
53669.80849  &      95.9  &    2.0  \\
54137.58873  &      31.0  &    1.5  \\
54143.51190  &      28.9  &    1.4  \\
54194.47989  &       8.6  &    1.5  \\
54315.91894  &     -38.8  &    2.3  \\
54384.87123  &     -60.3  &    3.2  \\
54431.68520  &     -75.4  &    1.9  \\
55563.54385  &     -63.1  &    1.0  \\
55564.57743  &     -66.2  &    0.8  \\
55903.70118  &      30.9  &    2.2  \\
56953.81794  &     -43.4  &    0.7  \\
56955.78182  &     -45.2  &    0.7  \\
56957.88054  &     -46.5  &    1.1  \\
56959.68147  &     -47.8  &    0.8  \\
\enddata
\end{deluxetable}

\begin{deluxetable}{lllll}
\tabletypesize{\scriptsize}
\tablecolumns{5}
\tablewidth{0pt}
\tablecaption{HD\,30177 Planetary System Parameters (all data) }
\tablehead{
\colhead{Parameter} & \multicolumn{2}{c}{Nominal Best Fit} & 
\multicolumn{2}{c}{Long-Period Solution} \\
\colhead{} & \colhead{HD\,30177b} & \colhead{HD\,30177c} & 
\colhead{HD\,30177b} & \colhead{HD\,30177c}
}
\startdata
\label{planetparams1}
Period (days) & 2532.5$\pm$10.6 & 6921$\pm$621 & 2520.6$\pm$8.9 & 11640$\pm$2432 \\
Eccentricity & 0.189$\pm$0.014 & 0.35$\pm$0.10 & 0.188$\pm$0.014 & 0.14$\pm$0.11 \\
$\omega$ (degrees) & 32$\pm$4 & 11$\pm$13 & 30$\pm$4 & 32$\pm$48 \\
$K$ (\ms) & 126.1$\pm$1.9 & 35.8$\pm$3.4 & 126.9$\pm$1.7 & 59.4$\pm$27.6 \\
$T_0$ (BJD-2400000) & 51428$\pm$26 & 51661$\pm$573 & 51434$\pm$24 & 48426$\pm$2978 \\
m sin $i$ (\Mjup) & 8.07$\pm$0.12 & 3.0$\pm$0.3 & 8.11$\pm$0.11 & 6.4$\pm$3.3 \\
$a$ (au) & 3.58$\pm$0.01 & 6.99$\pm$0.42 & 3.57$\pm$0.01 & 9.9$\pm$1.4 \\
\hline
RMS of fit (\ms) & 7.04  &   & 7.17  &   \\
$\chi^2_{\nu}$ (51 d.o.f.)  & 0.98 &  & 1.01  &  \\
\enddata
\end{deluxetable}


\begin{deluxetable}{lllll}
\tabletypesize{\scriptsize}
\tablecolumns{5}
\tablewidth{0pt}
\tablecaption{HD\,30177 Planetary System Parameters (outlier removed) }
\tablehead{
\colhead{Parameter} & \multicolumn{2}{c}{Nominal Best Fit} & 
\multicolumn{2}{c}{Long-Period Solution} \\
\colhead{} & \colhead{HD\,30177b} & \colhead{HD\,30177c} & 
\colhead{HD\,30177b} & \colhead{HD\,30177c}
}
\startdata
\label{planetparams2}
Period (days) & 2531.3$\pm$11.3 & 7601$\pm$1134 & 2524.4$\pm$9.8 & 11613$\pm$1837 \\
Eccentricity & 0.185$\pm$0.012 & 0.31$\pm$0.11 & 0.184$\pm$0.012 & 0.22$\pm$0.14 \\
$\omega$ (degrees) & 32$\pm$4 & 13$\pm$16 & 31$\pm$3 & 19$\pm$30 \\
$K$ (\ms) & 125.8$\pm$1.7 & 37.9$\pm$3.8 & 126.3$\pm$1.5 & 70.8$\pm$29.5 \\
$T_0$ (BJD-2400000) & 51430$\pm$27 & 52154$\pm$2009 & 51434$\pm$29 & 48973$\pm$1211 \\
m sin $i$ (\Mjup) & 8.06$\pm$0.11 & 3.32$\pm$0.45 & 8.08$\pm$0.10 & 7.6$\pm$3.1 \\
$a$ (au) & 3.58$\pm$0.01 & 7.45$\pm$0.75 & 3.58$\pm$0.01 & 9.89$\pm$1.04 \\
\hline
RMS of fit (\ms) & 5.98  &   & 6.01  &   \\
$\chi^2_{\nu}$ (50 d.o.f.)  & 0.76 &  & 0.77  &  \\
\enddata
\end{deluxetable}


\begin{figure}
\includegraphics[width=\textwidth]{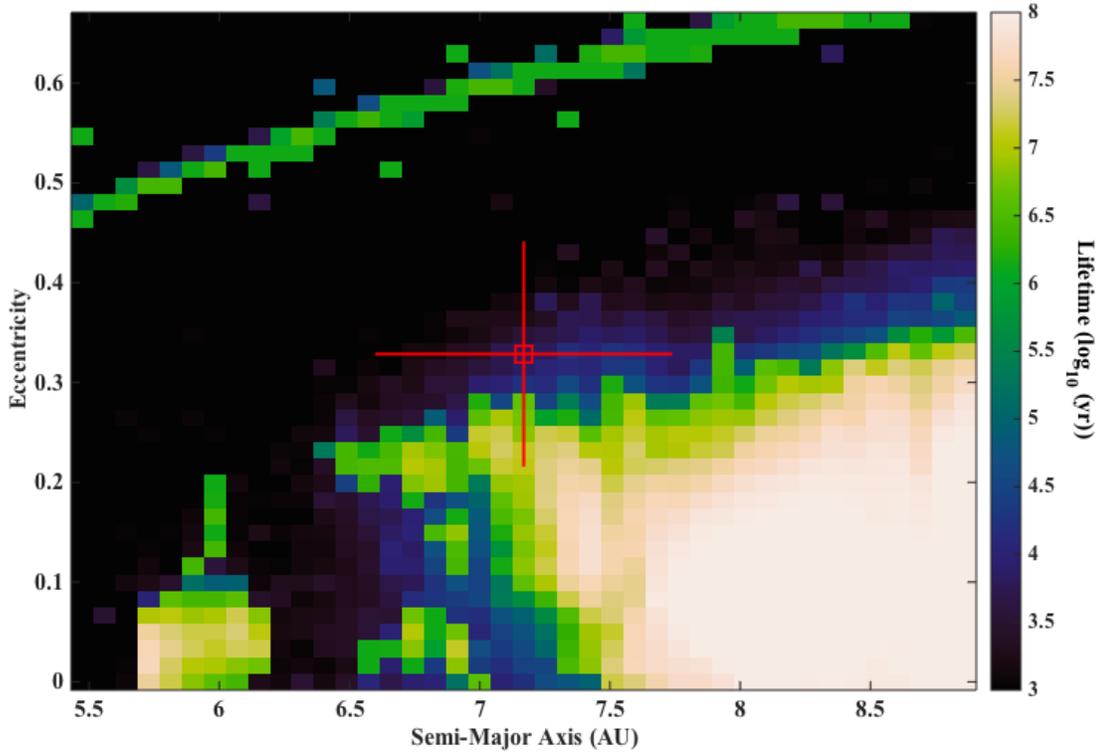}
\caption{The dynamical stability of the short-period solution for the 
orbit of HD 30177c, as a function of semi-major axis and eccentricity. 
The red box, to the centre of the plot, denotes the location of the 
best-fit solution, whilst the lines radiating from that point show the 
$1-\sigma$ uncertainties. It is immediately apparent that the best-fit 
solution lies in a region of significant dynamical instability.}
\label{ShortPeriodOrbit}
\end{figure}

\clearpage

\clearpage


\clearpage

\begin{figure}
\includegraphics[width=\textwidth]{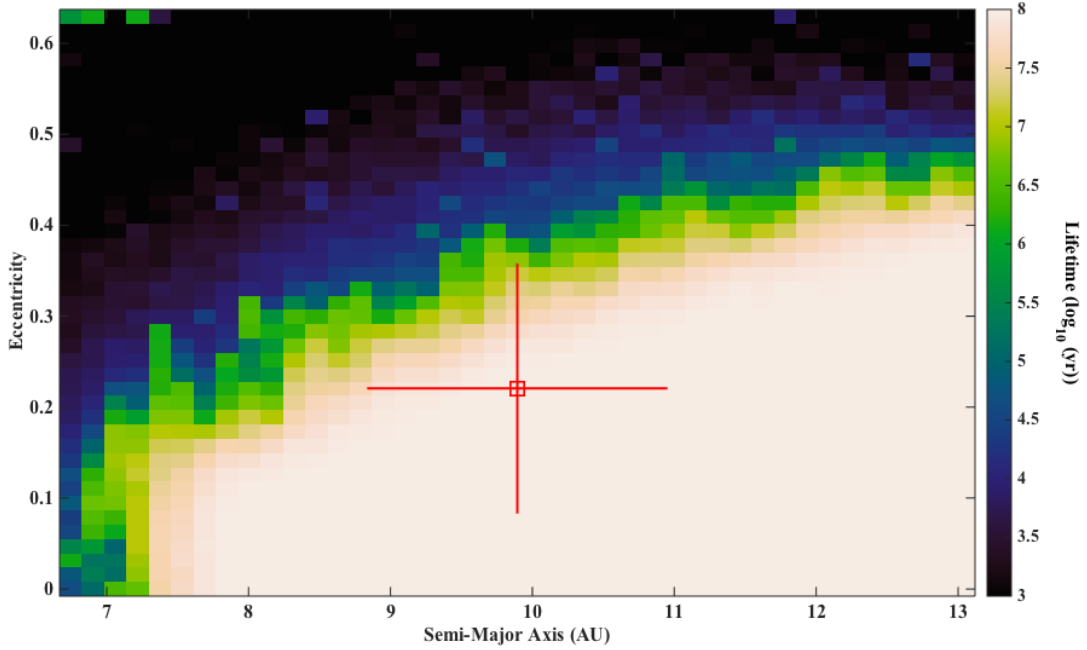}
\caption{The stability of the long-period solution for the orbit of 
HD\,30177c, as a function of semi-major axis and eccentricity. As with 
Figure ~\ref{ShortPeriodOrbit}, the red box marks the location of the 
best-fit solution, with the red lines radiating showing the $1-\sigma$ 
uncertainties. Unlike the short-period solution, the best-fit orbit now 
lies in a broad region of dynamical stability, with most solutions 
within $1-\sigma$ proving stable for the full 100 Myr of our 
integrations.}
\label{LongPeriodOrbit}
\end{figure}


\clearpage


\begin{thebibliography}{}

\bibitem[Adibekyan et al.(2012)]{adi12} Adibekyan, V.~Z., Sousa, S.~G., 
Santos, N.~C., et al.\ 2012, \aap, 545, A32

\bibitem[Anglada-Escud{\'e} \& Butler(2012)]{ang12} 
Anglada-Escud{\'e}, G., \& Butler, R.~P.\ 2012, \apjs, 200, 15

\bibitem[Borucki et al.(2011)]{borucki11} Borucki, W.~J., Koch, D.~G., 
Basri, G., et al.\ 2011, \apj, 736, 19

\bibitem[\protect\citeauthoryear{Butler et al.}{1996}]{BuMaWi96}
    Butler, R.~P., Marcy, G.~W., Williams, E., McCarthy, C.,
    Dosanjh, P., \& Vogt, S.~S. 1996, \pasp, {108}, 500

\bibitem[Butler et al.(2006)]{butler06} Butler, R.~P., Wright, J.~T., 
Marcy, G.~W., et al.\ 2006, \apj, 646, 505

\bibitem[Chambers(1999)]{Chambers99} 
Chambers, J.~E.\ 1999, \mnras, 304, 793 

\bibitem[Coughlin et al.(2016)]{coughlin16} Coughlin, J.~L., 
Mullally, F., Thompson, S.~E., et al.\ 2016, \apjs, 224, 12

\bibitem[Dumusque et al.(2011)]{dum11} Dumusque, X., 
Lovis, C., S{\'e}gransan, D., et al.\ 2011, \aap, 535, A55

\bibitem[Endl et al.(2016)]{endl16} Endl, M., Brugamyer, 
E.~J., Cochran, W.~D., et al.\ 2016, \apj, 818, 34

\bibitem[Fang \& Margot(2012)]{fang12} Fang, J., \& Margot, 
J.-L.\ 2012, \apj, 761, 92

\bibitem[Feng et al.(2015)]{feng15} Feng, Y.~K., Wright, J.~T., Nelson, 
B., et al.\ 2015, \apj, 800, 22

\bibitem[Fischer et al.(2016)]{fischer16} Fischer, D., 
Anglada-Escude, G., Arriagada, P., et al.\ 2016, arXiv:1602.07939

\bibitem[Franchini et al.(2014)]{franchini14} Franchini, M., 
Morossi, C., Marcantonio, P.~D., Malagnini, M.~L., \& Chavez, M.\ 2014, 
\mnras, 442, 220

\bibitem[Ghezzi et al.(2010)]{ghezzi10} Ghezzi, L., Cunha, K., Smith, 
V.~V., et al.\ 2010, \apj, 720, 1290

\bibitem[Han et al.(2014)]{han14} Han, E., Wang, S.~X., Wright, J.~T., 
et al.\ 2014, \pasp, 126, 827

\bibitem[H{\'e}brard et al.(2014)]{hebrard14} H{\'e}brard, 
{\'E}.~M., Donati, J.-F., Delfosse, X., et al.\ 2014, \mnras, 443, 2599

\bibitem[Hinse et al.(2014)]{hinse14} Hinse, T.~C., Horner, 
J., \& Wittenmyer, R.~A.\ 2014, Journal of Astronomy and Space Sciences, 
31, 187

\bibitem[Horner \& Jones(2008)]{hj08} Horner, J., \& Jones, B.~W.\ 2008, 
International Journal of Astrobiology, 7, 251

\bibitem[Horner \& Jones(2010b)]{hj10b} Horner, J., \& 
Jones, B.~W.\ 2010b, International Journal of Astrobiology, 9, 273

\bibitem[Horner et al.(2010)]{hjc10} Horner, J., Jones, B.~W., \& 
Chambers, J.\ 2010, International Journal of Astrobiology, 9, 1

\bibitem[Horner et al.(2011)]
{HUAqr} Horner, J., Marshall, J.~P., Wittenmyer, R.~A., \& Tinney, C.~G.\ 2011, \mnras, 416, L11 

\bibitem[Horner et al.(2013)]
{QSVir} Horner, J., Wittenmyer, R.~A., Hinse, T.~C., et al.\ 2013, \mnras, 435, 2033 

\bibitem[Horner et al.(2014)]{BDBlah} 
Horner, J., Wittenmyer, R.~A., Hinse, T.~C., \& Marshall, J.~P.\ 2014, \mnras, 439, 1176 

\bibitem[Houk \& Cowley(1975)]{houk75} Houk, N., \& Cowley, 
A.~P.\ 1975, University of Michigan Catalogue of two-dimensional 
spectral types for the HD stars.~Volume I.~Declinations -90\_ to 
-53\degrees., by Houk, N.; Cowley, A.~P..~ Ann Arbor, MI (USA): 
Department of Astronomy, University of Michigan, 19 + 452 p.,

\bibitem[Howard et al.(2010)]{howard10} Howard, A.~W., Johnson, J.~A., 
Marcy, G.~W., et al.\ 2010, \apj, 721, 1467

\bibitem[Jones et al.(2010)]{jones10} Jones, H.~R.~A., Butler, R.~P., 
Tinney, C.~G., et al.\ 2010, \mnras, 403, 1703

\bibitem[Kennedy et al.(2013)]{kennedy13} Kennedy, G.~M., 
Wyatt, M.~C., Bryden, G., Wittenmyer, R., \& Sibthorpe, B.\ 2013, 
\mnras, 436, 898

\bibitem[Lewis et al.(2013)]{lewis13} Lewis, A.~R., Quinn, T., \& Kaib, 
N.~A.\ 2013, \aj, 146, 16

\bibitem[Lovis et al.(2011)]{lovis11} Lovis, C., Dumusque, 
X., Santos, N.~C., et al.\ 2011, arXiv:1107.5325

\bibitem[Marmier et al.(2013)]{marmier13} Marmier, M., S{\'e}gransan, 
D., Udry, S., et al.\ 2013, \aap, 551, A90

\bibitem[Marshall et al.(2010)]{MarshallHR8799} 
Marshall, J., Horner, J., \& Carter, A.\ 2010, International Journal of Astrobiology, 9, 259 

\bibitem[Meschiari et al.(2009)]{mes09} Meschiari, S., Wolf, A.~S., 
Rivera, E., et al.\ 2009, PASP, 121, 1016

\bibitem[Moutou et al.(2015)]{moutou15} Moutou, C., Lo Curto, G., Mayor, 
M., et al.\ 2015, \aap, 576, A48

\bibitem[O'Toole et al.(2009)]{otoole09} O'Toole, S.~J., 
Tinney, C.~G., Jones, H.~R.~A., et al.\ 2009, \mnras, 392, 641

\bibitem[Robertson et al.(2012)]{HD204313} 
Robertson, P., Horner, J., Wittenmyer, R.~A., et al.\ 2012, \apj, 754, 50 

\bibitem[Robertson et al.(2012)]{texas1} Robertson, P., Endl, M., 
Cochran, W.~D., et al.\ 2012, \apj, 749, 39

\bibitem[Rowan et al.(2016)]{rowan16} Rowan, D., Meschiari, S., 
Laughlin, G., et al.\ 2016, \apj, 817, 104

\bibitem[Rowe et al.(2014)]{rowe14} Rowe, J.~F., Bryson, S.~T., Marcy, 
G.~W., et al.\ 2014, \apj, 784, 45

\bibitem[Santos et al.(2013)]{santos13} Santos, N.~C., Sousa, 
S.~G., Mortier, A., et al.\ 2013, \aap, 556, A150

\bibitem[Sousa et al.(2011)]{sousa11} Sousa, S.~G., Santos, N.~C., 
Israelian, G., Mayor, M., \& Udry, S.\ 2011, \aap, 533, A141

\bibitem[Takeda et al.(2007)]{takeda07} Takeda, G., Ford, E.~B., Sills, 
A., et al.\ 2007, \apjs, 168, 297

\bibitem[Tinney et al.(2001)]{tinney01} Tinney, C.~G., Butler, R.~P., 
Marcy, G.~W., et al.\ 2001, \apj, 551, 507

\bibitem[Tinney et al.(2003)]{tinney03} Tinney, C.~G., Butler, R.~P., 
Marcy, G.~W., et al.\ 2003, \apj, 587, 423

\bibitem[Tinney et al.(2011)]{tinney11} Tinney, C.~G., 
Wittenmyer, R.~A., Butler, R.~P., et al.\ 2011, \apj, 732, 31

\bibitem[\protect\citeauthoryear{Valenti et~al.} {1995}]{val:95}
    Valenti, J.~A., Butler, R.~P. \& Marcy, G.~W. 1995,
    \newblock { PASP, } {107}, 966.

\bibitem[van Leeuwen(2007)]{vl07} van Leeuwen, F.\ 2007, \aap, 474, 653

\bibitem[Wittenmyer et al.(2011)]{jupiters} Wittenmyer, 
R.~A., Tinney, C.~G., O'Toole, S.~J., et al.\ 2011, \apj, 727, 102

\bibitem[Wittenmyer et al.(2012a)]{142paper} Wittenmyer, R.~A., Horner, 
J., Tuomi, M., et al.\ 2012a, \apj, 753, 169

\bibitem[Wittenmyer et al.(2012b)]{HUAqr2} Wittenmyer, R.~A., Horner, 
J., Marshall, J.~P., Butters, O.~W., \& Tinney, C.~G.\ 2012b, \mnras, 
419, 3258

\bibitem[Wittenmyer et al.(2012c)]{24Sex} 
Wittenmyer, R.~A., Horner, J., \& Tinney, C.~G.\ 2012, \apj, 761, 165 

\bibitem[Wittenmyer et al.(2013)]{songhu} Wittenmyer, 
R.~A., Wang, S., Horner, J., et al.\ 2013, \apjs, 208, 2

\bibitem[Wittenmyer et al.(2014a)]{2jupiters} Wittenmyer, R.~A., Horner, 
J., Tinney, C.~G., et al.\ 2014a, \apj, 783, 103

\bibitem[Wittenmyer et al.(2014b)]{gj832} Wittenmyer, 
R.~A., Tuomi, M., Butler, R.~P., et al.\ 2014b, \apj, 791, 114

\bibitem[Wittenmyer et al.(2014c)]{HD73526} 
Wittenmyer, R.~A., Tan, X., Lee, M.~H., et al.\ 2014c, \apj, 780, 140 

\bibitem[Wittenmyer et al.(2016a)]{ppps3} Wittenmyer, 
R.~A., Butler, R.~P., Wang, L., et al.\ 2016a, \mnras, 455, 1398

\bibitem[Wittenmyer et al.(2016b)]{newjupiters} Wittenmyer, R.~A., 
Butler, R.~P., Tinney, C.~G., et al.\ 2016b, \apj, 819, 28

\bibitem[Wright et al.(2011)]{wright11} Wright, J.~T., Fakhouri, O.,
Marcy, G.~W., et al.\ 2011, \pasp, 123, 412



\end{thebibliography}
\end{document}